\documentclass[a4paper,fleqn,usenatbib]{mnras}

\usepackage{newtxtext,newtxmath}

\usepackage[T1]{fontenc}
\usepackage{ae,aecompl}
\usepackage{graphicx}
\usepackage{amsmath}
\usepackage{anyfontsize}
\usepackage[utf8]{inputenc}

\usepackage{etoolbox}
\defcitealias{Dullemond2018}{D18}

\title[The efficiency of dust trapping]{The efficiency of dust trapping in ringed proto-planetary discs}

\author[G. P. Rosotti et al.]{Giovanni~P.~Rosotti$^{1}$\thanks{rosotti@strw.leidenuniv.nl},
 Richard Teague$^{2,3}$, Cornelis Dullemond$^{4}$, Richard A. Booth$^{5}$, \newauthor Cathie Clarke$^{5}$\\
$^{1}$Leiden Observatory, Leiden University, P.O.~Box 9513, NL-2300~RA Leiden, the Netherlands\\
$^{2}$Department of Astronomy, University of Michigan, 1085 South University Avenue, Ann Arbor, MI 48109, USA\\
$^{3}$Center for Astrophysics | Harvard \& Smithsonian, 60 Garden Street, Cambridge, MA 02138 USA\\
$^{4}$Zentrum für Astronomie, Institut für Theoretische Astrophysik, Universität Heidelberg, Albert-Ueberle-Str. 2, 69120 Heidelberg, Germany\\
$^{5}$Institute of Astronomy, Madingley Road, Cambridge, CB3 0HA, UK
\vspace{-3mm}
}

\date{Accepted 2020 April 22. Received 2020 April 15; in original form 2020 February 4}\vspace{-2mm}

\pubyear{2020}

\begin{document}
\label{firstpage}
\pagerange{\pageref{firstpage}--\pageref{lastpage}}
\maketitle

\begin{abstract}
When imaged at high-resolution, many proto-planetary discs show gaps and rings in their dust sub-mm continuum emission profile. These structures are widely considered to originate from local maxima in the gas pressure profile. The properties of the underlying gas structures are however unknown. In this paper we present a method to measure the dust-gas coupling $\alpha/St$ and the width of the gas pressure bumps affecting the dust distribution, applying high-precision techniques to extract the gas rotation curve from emission lines data-cubes. As a proof-of-concept, we then apply the method to two discs with prominent sub-structure, HD163296 and AS 209. We find that in all cases the gas structures are larger than in the dust, confirming that the rings are pressure traps. Although the grains are sufficiently decoupled from the gas to be radially concentrated, we find that the degree of coupling of the dust is relatively good ($\alpha/St \sim 0.1$). We can therefore reject scenarios in which the disc turbulence is very low and the dust has grown significantly. If we further assume that the dust grain sizes are set by turbulent fragmentation, we find high values of the $\alpha$ turbulent parameter ($\alpha \sim 10^{-2}$). Alternatively, solutions with smaller turbulence are still compatible with our analysis if another process is limiting grain growth. For HD163296, recent measurements of the disc mass suggest that this is the case if the grain size is 1mm. Future constraints on the dust spectral indices will help to discriminate between the two alternatives.
\end{abstract}


%%%%%%%%%%%%%%%%% BODY OF PAPER %%%%%%%%%%%%%%%%%%

\section{Introduction}
The Atacama Large Millimeter/submillimeter Array (ALMA) is revolutionising our understanding of proto-planetary discs thanks to its unprecedented angular resolution. When imaged at high resolution, most (though not all, \citealt{Facchini2019,Long2019}) discs show a rich morphology of structures, in terms of crescents \citep{vanderMarel2013}, spirals \citep{Perez2016} and rings \citep{2015ApJ...808L...3A,vanderPlas2017,Fedele2017,Fedele2018,Dipierro2018,Clarke2018}. This latter category in particular is the one occurring most frequently, as shown spectacularly by the high-resolution DSHARP campaign \citep{Andrews2018} and by other efforts with large disc samples \citep{Long2018,vanderMarel2019}.

These rings are interesting for many reasons. Firstly, they are thought to be dust traps, where the dust stops drifting towards the star and accumulates. In this sense, they could be the solution to the long-standing problem of how to reduce the importance of radial drift, which if unimpeded would deplete discs on a very short timescale \citep{TakeuchiLin2005,Brauer2007}, leaving little solid mass to form the rocky planetary cores \citep{Greaves2010,Manara2018,Rosotti2019}. Secondly, the most likely interpretation for the origin of these rings is that a population of young planets is \textit{already} present at these early stages; the rings are therefore a tool to study the masses and locations of these young planets \citep{Rosotti2016,Bae2018,Zhang2018,Lodato2019}.

Independently from their formation mechanisms, there is another sense in which the commonly imaged rings are important: they provide us with new windows to probe disc physics. %, measuring disc parameters that are otherwise impossible or very difficult to constrain. 
One example is the magnitude of the turbulence, another long standing problem in planet formation \citep{1974MNRAS.168..603L}, typically parametrised through the dimensionless $\alpha$ parameter \citep{1973A&A....24..337S}. The amount of turbulence is a crucial parameter regulating, just to name a few examples, the efficiency of gas accretion onto the star and forming planets \citep{Bodenheimer2013}, how discs responds to planets \citep{2012ARA&A..50..211K,Zhang2018}, the vertical mixing of molecular species \citep{SemenovWiebe2011}, the importance of fragmentation for dust evolution \citep{OrmelCuzzi2007,Birnstiel2012}, and many other processes. Because turbulence in proto-planetary discs is expected to be highly sub-sonic, degeneracies with the disc temperature mean however that the turbulence is proving very difficult to constrain directly \citep{Teague2016,Flaherty2017,Teague2018CS} from broadening of emission lines. This is where rings come to the rescue, since the dust is also subject to turbulence and can also be employed as an observational tracer of turbulence. To be more precise, in this way it is only possible to measure $\alpha/St$ rather than $\alpha$, where $St$ is the so-called Stokes parameter quantifying the aerodynamic coupling between gas and dust. \citet{Pinte2016} showed that turbulence in the \textit{vertical} direction can be measured by quantifying the degree of smoothing of the emission profile along the disc semi-minor axis. With a complementary method, \citet[][hereafter D18]{Dullemond2018} used the radial width of dust rings to put constraints on the turbulence in the \textit{radial} direction. Unfortunately, with their methodology a turbulence measurement requires comparing the radial width of features in the dust and gas surface densities, but only data regarding the former were available. Therefore, they were able only to identify a \textit{range} of permitted values and not to measure the \textit{value} of $\alpha/St$.% they employed only dust data Ultimately, however, t because they only employed dust data.

Thankfully, there is a way forward to improve on the analysis of \citetalias{Dullemond2018}. In addition to the continuum emission, ALMA is also revolutionising our view of the gas disc. Thanks to the combination of ALMA extreme sensitivity and spatial resolution, plus new techniques developed to make use of these innovative data, the gas rotational velocity can now be studied with high precision. As highlighted by a few spectacular examples \citep{Teague2018,Pinte2018,Teague2019,Dullemond2019}, there is now a growing realisation that most discs are not in perfect Keplerian rotation, with deviations amounting to a few percent. In our current understanding of disc dynamics, these deviations in the gas velocity are the very reason why we observe rings in the dust distribution \citep{1972fpp..conf..211W}.

Applying these techniques to the DSHARP data opens up the possibility of measuring the turbulence by combining information about the dust and the gas. In addition, measuring the width of gas structures directly confirms that these structures are dust traps if the gas width is larger than the dust width \citepalias[see discussion in][]{Dullemond2018}. Performing these measurements is the goal of this paper. As a proof-of-concept, we focus here on the two discs with most prominent structures in gas and continuum, namely HD163296 and AS 209.

The paper is structured as follows. Section \ref{sec:methods} introduces the method we use to measure the dust-gas coupling and the width of gas structures. We then present our results and discuss possible caveats in section \ref{sec:results}. Finally, we draw our conclusions in section \ref{sec:discussion_conclusion}.

\section{Methods}
\label{sec:methods}

Our analysis is based on the publicly available data from the DSHARP ALMA large programme\footnote{\url{https://almascience.eso.org/almadata/lp/DSHARP/}}, focusing on the discs of HD~163296 and AS~209 \citep{Andrews2018, Isella2018, Guzman2018}. Our goal is to measure the dust-gas coupling $\alpha/St$. The new aspect of this paper is that we use the $^{12}$CO data-cubes to measure the slope of the deviation from Keplerian rotation of the gas in the proximity of the continuum peaks. As we will show, in combination with the width of the dust rings, this can be used to yield a measurement of $\alpha/St$. Additionally, from the same data, we can also measure the width of the rings in the gas distribution.% With the methodology of \citet{Dullemond2018}, we use the continuum images to measure the width of dust structures.  %We detail below our procedure.

%We follow \citet{Teague2018AS209} to extract the gas rotation curves from the DSHARP cubes. From the channel maps, we construct a projected velocity map (note that this is not the first moment map) following the method of \citet{bettermoments}, which allows us to discriminate between the emission coming from the disc near and far sides. The individual channel maps are also used to fit an emission surface using the method of \citet{Pinte2018}. Armed with this knowledge, we deproject the data to polar coordinates taking into account of the finite height of the emission surface. Finally, we do some black magic with Gaussian Processing to derive the azimuthally averaged rotation curve. \gr{richard, please write some sensible text about the GP}

\subsection{Calculating the rotation curve}

To calculate the rotation curve we follow the method described in \citet{Teague2018AS209} which is broken into two aspects: the measurement of the $^{12}$CO emission surface in order to correctly deproject the data into annuli, and secondly the inference of $v_{\phi}$ in each annulus.

Measuring the emission surface is done by fitting the map of line centres, made using \texttt{bettermoments} \citep{bettermoments}, which fits a quadratic curve to the pixel of peak intensity and two neighbouring pixels, with a Keplerian rotation pattern including a correction for the 3D geometry of the disk, as described in \citet{Keppler2019} using the \texttt{eddy} Python package \citep{eddy}. As AS~209 has cloud contaminated regions, we additionally use the method described in \citet{Pinte2018IMLup} which is less sensitive to cloud-contamination to verify the emission surfaces we obtain. We find excellent agreement with previous determinations in these sources \citep{Teague2018AS209, Isella2018}.% \rt{Should we include the actual $z(r)$ profiles?} \gr{I think it's a good idea. They can probably go in an appendix? (since we don't have many already :) )}.

Using these emission surfaces we deproject the data into disk-centric coordinates, $(r,\, \phi)$, and divide them into annuli with a width of 1/4 of the beam major axis. We stress that this binning does not remove the spatial correlation between nearby annuli, however minimises the impact of Keplerian shear across the beam when measuring $v_{\phi}$. Within each annulus the projected component of $v_{\phi}$ is assumed to vary as a function of azimuth, $v_{\phi,\,{\rm proj}} = v_{\phi} \cdot \cos \phi \cdot \sin i$, where $\phi$ is measured from the red-shifted major axis of the disk and $i$ is the disk inclination. Using \texttt{eddy} \citep{eddy}, $v_{\phi}$ is inferred by finding the value of $v_{\phi}$ which allows for the spectra to be shifted back to a common line center (the systemic velocity). More details of the exact fitting procedure can be found in \citet{Teague2018AS209}. This procedure is repeated for each annulus, yielding $v_{\phi}$ as a function of radius. In order to measure the deviation $\delta v_{\phi}=v_{\phi}-v_K$ from the Keplerian value $v_K$, we fit the $v_{\phi}$ profiles with a double power-law profile. While a single power-law would be sufficient for a purely Keplerian rotation profile, the inclusion of radial pressure gradients and changes in the emission height with radius result in systematic deviations from a pure $v_K$ Keplerian profile. An important fact to stress is that in reality we do not know the true Keplerian value, because we do not know precisely enough the stellar mass. As a surrogate, we employ the deviation from the double power-law fit, implying that there might be a constant, unknown offset (with magnitude of a few per cent) between our $\delta v_\mathrm{\phi}$ and the deviation from Keplerian - but for simplicity, in the rest of the paper we will often refer to $\delta v_\mathrm{\phi}$ as "deviation from Keplerian". Our analysis is not affected by this offset since we will show that it relies on the derivative.%\rt{Might need to think about how to better justify this.}

\subsection{Using the rotation curve to measure \texorpdfstring{$\alpha/St$}{α/St} and gas widths}

\begin{figure}
    \centering
    \includegraphics[width=\columnwidth]{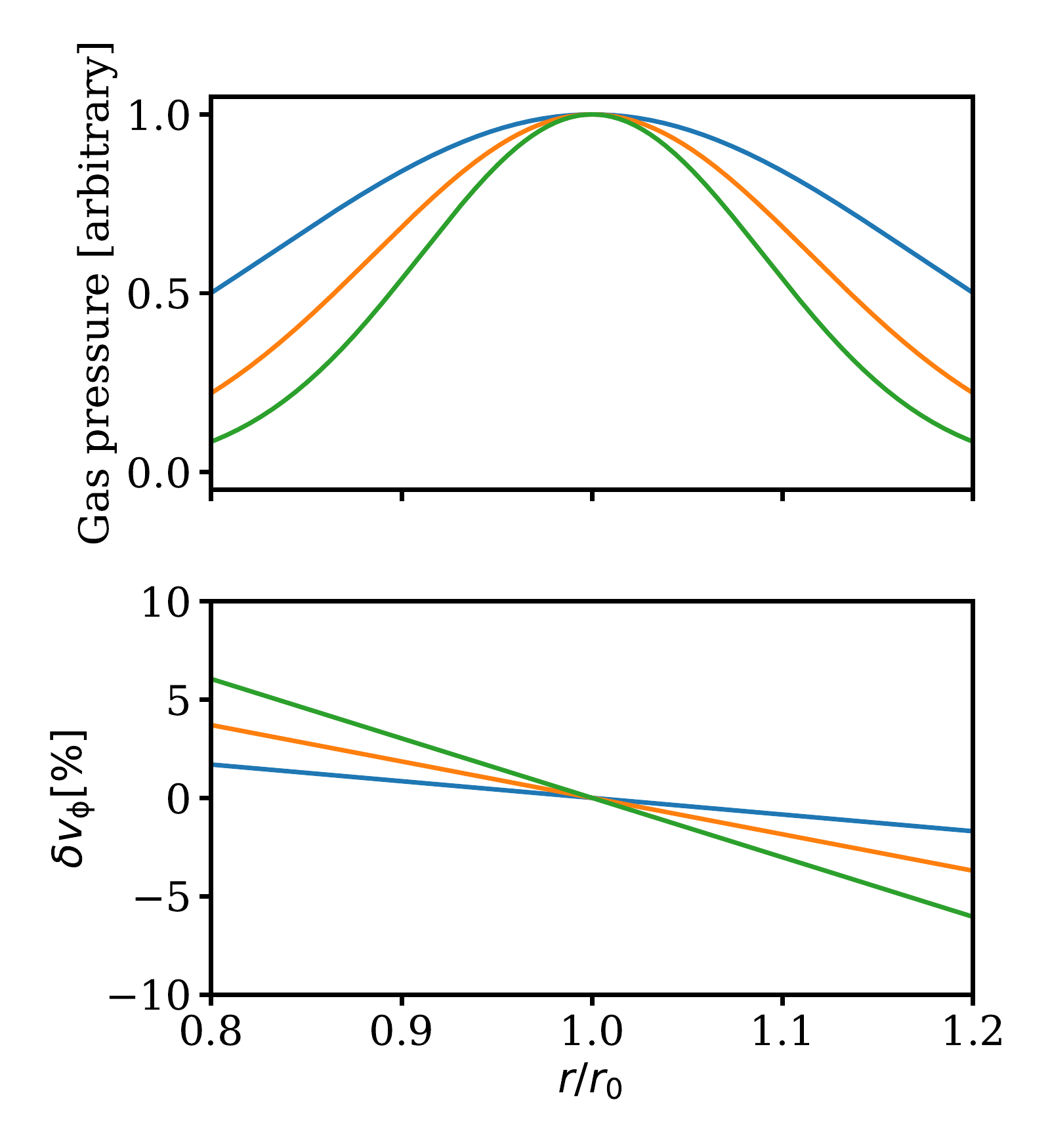}
    \caption{Graphical illustration of the method we use in this paper (see \autoref{eq:width_gas_main} and appendix \ref{sec:rot_width}) to measure the gas width: the width of a Gaussian surface density profile (top panel) is linked to the steepness of the deviation of the rotation curve from Keplerian (bottom panel). %In the bottom panel we plot the derivative of the curve in the middple panel. The dashed lines are for the analytical approximation we employ in this paper (see main text and appendix \ref{sec:rot_width}), while the solid lines are the exact solution; there is a good match between the two.
    }
    \label{fig:gaussian_bump}
\end{figure}

By studying $\delta v_\phi$, we can now determine $\alpha/St$. Assuming that the disc is razor-thin, we show in appendix \ref{sec:rot_width} that, to first order in $r-r_0$, the dust surface density is a Gaussian with width $w_d$. The following expression (see \autoref{eq:alpha_St}) links the width $w_d$ with the dust-gas coupling and the observables:
\begin{equation}
\frac{\alpha}{St} = - \frac{2 w_d^2}{r_0} \frac{v^2_k}{c_s^2} \frac{\mathrm{d}}{\mathrm{d}r} \left( \frac{\delta v_\phi}{v_k}\right).
\label{eq:alpha_St_main}
\end{equation}

As we derive in appendix \ref{sec:width_gas}, the same observables we use to measure $\alpha/St$ can also be used to measure the width of the gas rings $w_g$, using the following expression (see \autoref{eq:width_gas}):
\begin{equation}
w_g=\sqrt{-\frac{1}{2}\frac{c_s^2}{v_K^2}r_0\left[\frac{d}{dr}\left(\frac{\delta v_\phi}{v_K}\right)\right]^{-1}}.
\label{eq:width_gas_main}
\end{equation}
\autoref{fig:gaussian_bump} illustrates graphically this method, showing how Gaussians of different width produce a different gradient in the deviation from Keplerian rotation.

In the expressions above, $r_0$ is trivially obtained as the location of the dust ring. We already discussed in the previous section how we derive the rotation curve and we discuss more in detail in section \ref{sec:derived_width} how we measure the slope. For what concerns the dust width $w_d$, we measure it from the continuum images as in \citetalias{Dullemond2018}. We discuss in the next paragraph the last parameter, the gas temperature.

%Once this is known, we can measure the dust diffusivity. 

The razor-thin model should be considered only as pedagogical since it is very well known that the CO emission comes from an elevated surface. Therefore, a proper modelling should take into account the disc vertical structure. We show in appendix \ref{sec:vertical_structure} that in practice this can be accounted for using the gas temperature at the emitting layer, instead of the midplane temperature, for computing $c_s^2$ in \autoref{eq:alpha_St_main} and \autoref{eq:width_gas_main}. %The additional terms can be neglected, %by employing the gas temperature at the emitting layer and additional terms can be neglected, as long as the temperature structure is a smooth function of radius. 
At the emitting layer, the temperature can be estimated with high precision from the $^{12}$CO data using the peak brightness temperature given the high optical depth of $^{12}$CO; we therefore use directly these values from the data.% \gr{This statement I have written is only partially true. Looking at fig. 7 of isella et al 2018 and fig. 5 of guzman et al 2018, there are structures in the CO emission. But they are small, and they both comment that it could just be the continuum becoming partially optically thick close to the ring. Maybe I should make these statements a little bit less strong?}

%Our model is by necessity simplistic and it is worth discussing the effect of the assumptions we have taken concerning the gas surface density and the disc vertical structure. We discuss below how they do not significantly affect our conclusions and how our model is therefore effective in catching the basic physics of dust trapping.

\section{Results}
\label{sec:results}

\begin{figure*}
    \centering
    \includegraphics[width=\columnwidth]{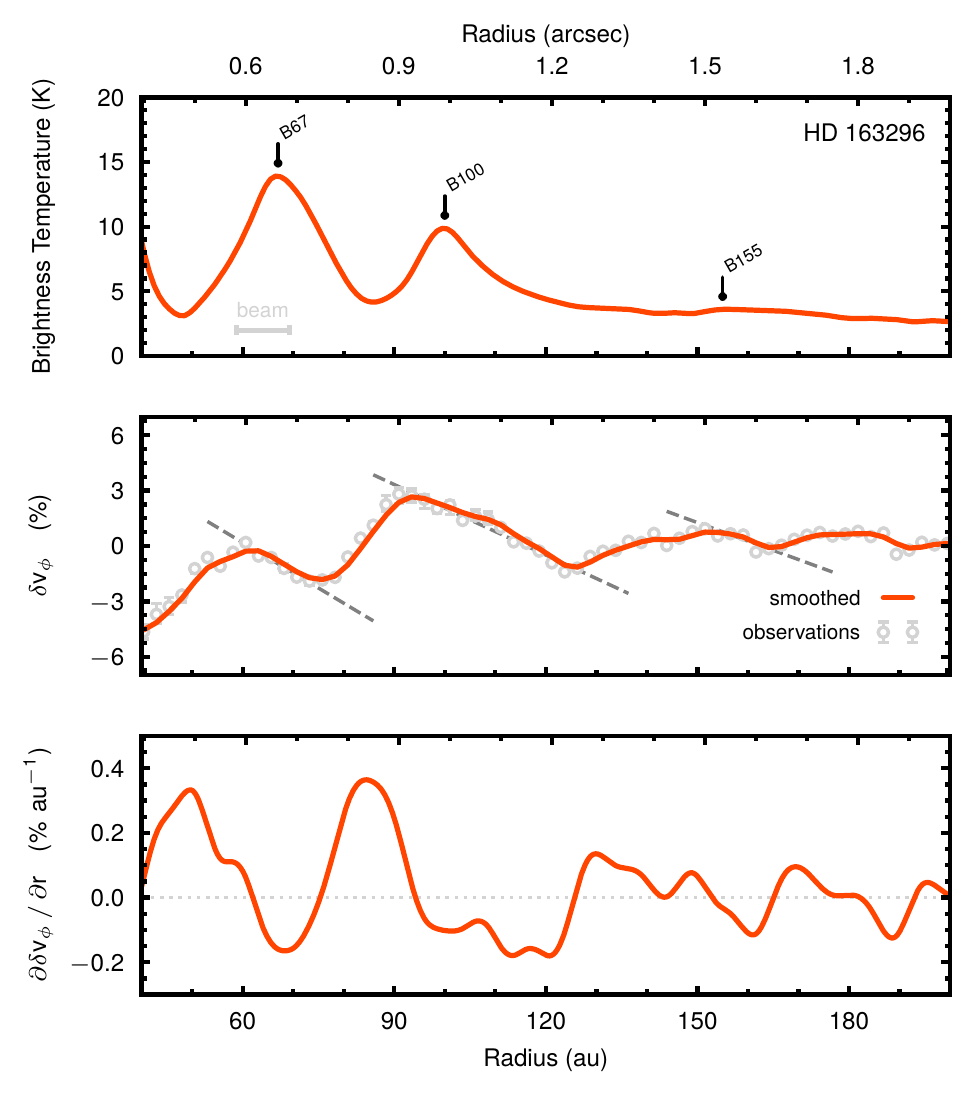}
    \quad
    \includegraphics[width=\columnwidth]{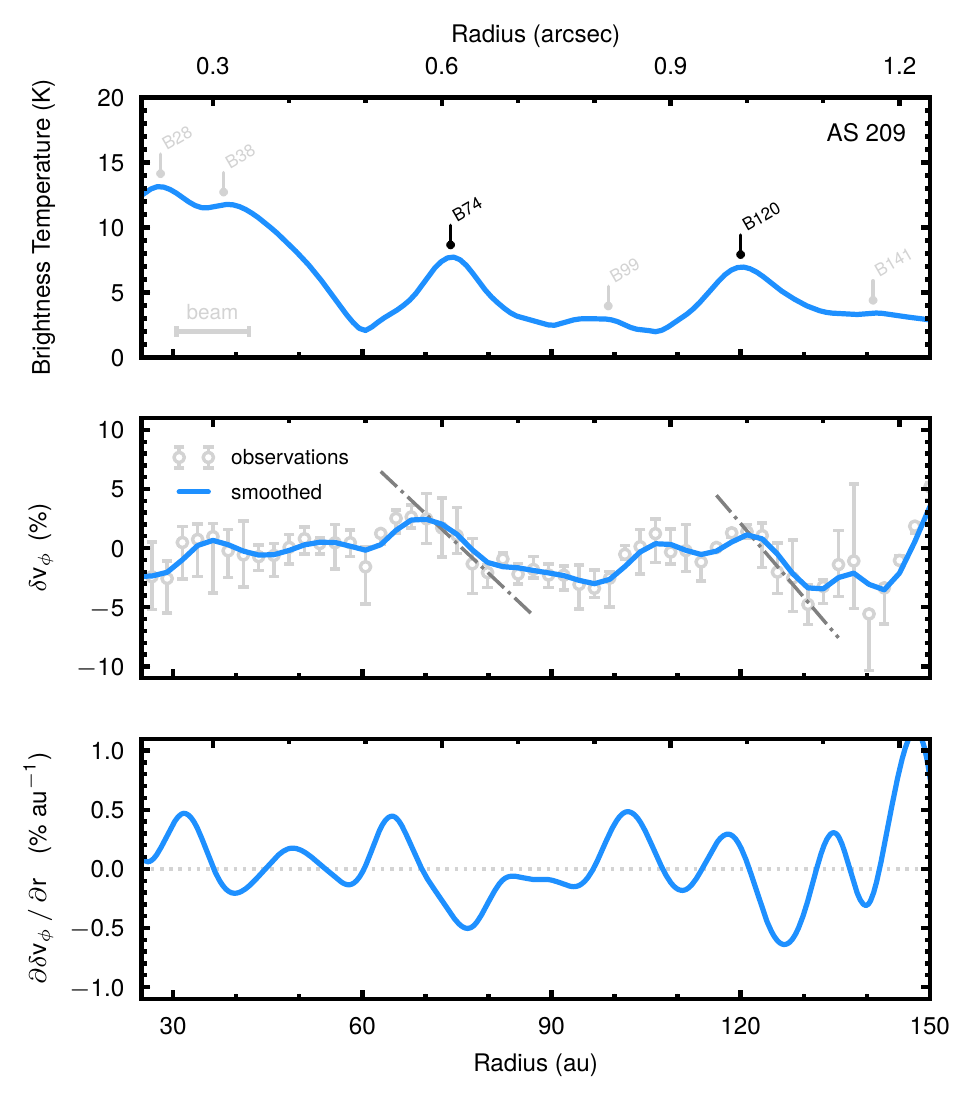}
    \caption{Data for HD163296 and AS 209, left and right, respectively. \textit{Top panel}: continuum emission profiles. We marked the location of the continuum peaks, using the notation of \citet{Huang2018}. The FWHM of the synthesized beams are shown in the bottom left of each panel: 104~mas and 94~mas, respectively. \textit{Middle panel}: rotation curves derived from the observations. We marked with the grey dashed lines the linear fits in the vicinity of the continuum peaks. \textit{Bottom panel}: derivative of $\delta v_\phi$.}
    \label{fig:vrot_observations}
\end{figure*}

% The average gradients (in units of [%/au]) over the same fit range are:

%     HD 163296
%     B67 = -0.14 +\- 0.04
%     B100 = -0.014 +\- 0.04
%     B155 = -0.07 +\- 0.04

%     AS 209
%     B74 = -0.31 +\- 0.15
%     B120 = -0.35 +\- 0.24
    
% where the uncertainties are the standard deviation. All are consistent within 1 sigma of the linear fit.

\begin{table*}
\caption{Values derived from the observations for the five pressure traps analysed in this paper with 1$\sigma$ uncertainties.}
\label{tab:results}
%\centering
\begin{tabular}{lcccccccc} \hline\hline
 & (1) & (2) & (3) & (4) & (5) & (6) & (7) & (8) \\
Ring & $\partial \delta v_{\phi} / \partial r$ & $T_B$ & $v_{\rm kep}$ & $w_{\rm dust}$ & $\alpha / {\rm St}$ & $w_{\rm gas}$ & $\Delta r$ & $\displaystyle \frac{\Delta v_{\phi}} { (c_s/v_K)^2}$   \\
 & ($\%~{\rm au}^{-1}$) & (K) &  (${\rm m\,s^{-1}}$) & (${\rm au}$) &  & (${\rm au}$) & (${\rm au}$)\\ \hline
\multicolumn{9}{c}{\footnotesize HD~163296} \\ \hline
B67     & $-0.20 \pm 0.02$  & $81.5 \pm 8.2$ & $4784 \pm 4$ & $6.85 \pm 0.03$ & $0.23 \pm 0.03$ &$14.4 \pm 1.0$ & 7 & 1.1 \\ 
B100    & $-0.15 \pm 0.01$ & $71.7 \pm 6.2$  & $3932 \pm 2$ & $4.66 \pm 0.08$ & $0.04 \pm 0.01$ &$23.2 \pm 1.3$ & 15 & 1.4\\
B155    & $-0.12 \pm 0.02$ & $68.3 \pm 5.1$  & $3186 \pm 1$ & $7.25 \pm 1.77$ & $0.04 \pm 0.02$ &$34.8 \pm 2.7$& 0.02 & 0.4 \\
\hline
\multicolumn{9}{c}{\footnotesize AS~209} \\ \hline           
B74     & $-0.50 \pm 0.05$ & $41.6 \pm 4.5$ & $4092 \pm 7$ & $3.39 \pm 0.06$ & $0.18 \pm 0.04$ &  $8.0 \pm 0.6$& 5 & 2.9\\
B120    & $-0.62 \pm 0.06$ & $37.0 \pm 2.8$  & $3146 \pm 4$ & $4.12 \pm 0.07$ & $0.13 \pm 0.02$ &$11.2 \pm 0.7$& 10 & 4.8\\
\hline
%(${\rm m\,s^{-1}}$) & $533 \pm 27$ &  & $500 \pm 22$ & $488 \pm 18$ & $381 \pm 20$ & $359 \pm 14$
\end{tabular}
\par \textit{Notes} All quantities are evaluated at the location of each dust ring. (1) Slope of the deviation of the rotation curve from Keplerian (2) Brightness temperature, used to estimate the temperature at the emitting layer (3) Keplerian velocity (4) Width of the dust ring (5) Value of $\alpha/St$ computed using \autoref{eq:alpha_St_main} (6) Width of the gas ring computed using \autoref{eq:width_gas_main}. (7) Radial extent over which the deviation from Keplerian has a negative slope (8) See \ref{sec:caveats}.
\end{table*}

\subsection{Derived values}
\label{sec:derived_width}

We show in the middle panel of \autoref{fig:vrot_observations} the rotation curves extracted from the data. For comparison we show also the continuum profiles on the top panels. We note that in the vicinity of the continuum peaks $\delta v_\phi$ \textit{decreases}, as expected in the case of a pressure maximum. The curves are also reasonably well described by a constant slope, with the most spectacular example in the vicinity of the B100 peak of HD~163296, with a relatively extended radial range. Overall there is therefore reasonable agreement between the dust structure and the rotation curve. That being said, we note that in the B100 peak of HD~163296 and in the B120 peak of AS~209 the region with a decreasing $\delta v_\phi$ is not symmetrical with respect to the continuum peak, as one might instead expect. We discuss possible explanations for this in section \ref{sec:caveats}.

To measure the slope, we find that the raw derivative (bottom panel) can be relatively noisy, given the data spatial resolution and signal to noise; we thus perform a linear fit which is more robust towards the noise and correctly accounts for uncertainties. We list in \autoref{tab:results} the measured gradients $\partial \delta v_{\phi} / \partial r$.

In \autoref{tab:results} we list also the gas temperature $T_B$ close to the peak, measured using the peak brightness temperature (see section \ref{sec:methods}), and the Keplerian velocity $v_{\rm kep}$ (estimated via the double power-law fit). To compute the sound speed, we assume a mean molecular weight $\mu = 2.37$. We also list the dust widths $w_d$ measured by fitting Gaussians in the azimuthally averaged continuum profiles \citep{Andrews2018, Huang2018}, which agree in all cases with those measured by \citetalias{Dullemond2018}, except for B155 which was not analysed by them \citepalias[in this case we fitted the continuum emission between 150 and 158~au following the procedure in][]{Dullemond2018}. As already argued by \citetalias{Dullemond2018}, the finite radial extent of the dust rings implies that some mechanism is stirring the dust in the radial direction.

\begin{figure*}
\includegraphics[width=\columnwidth]{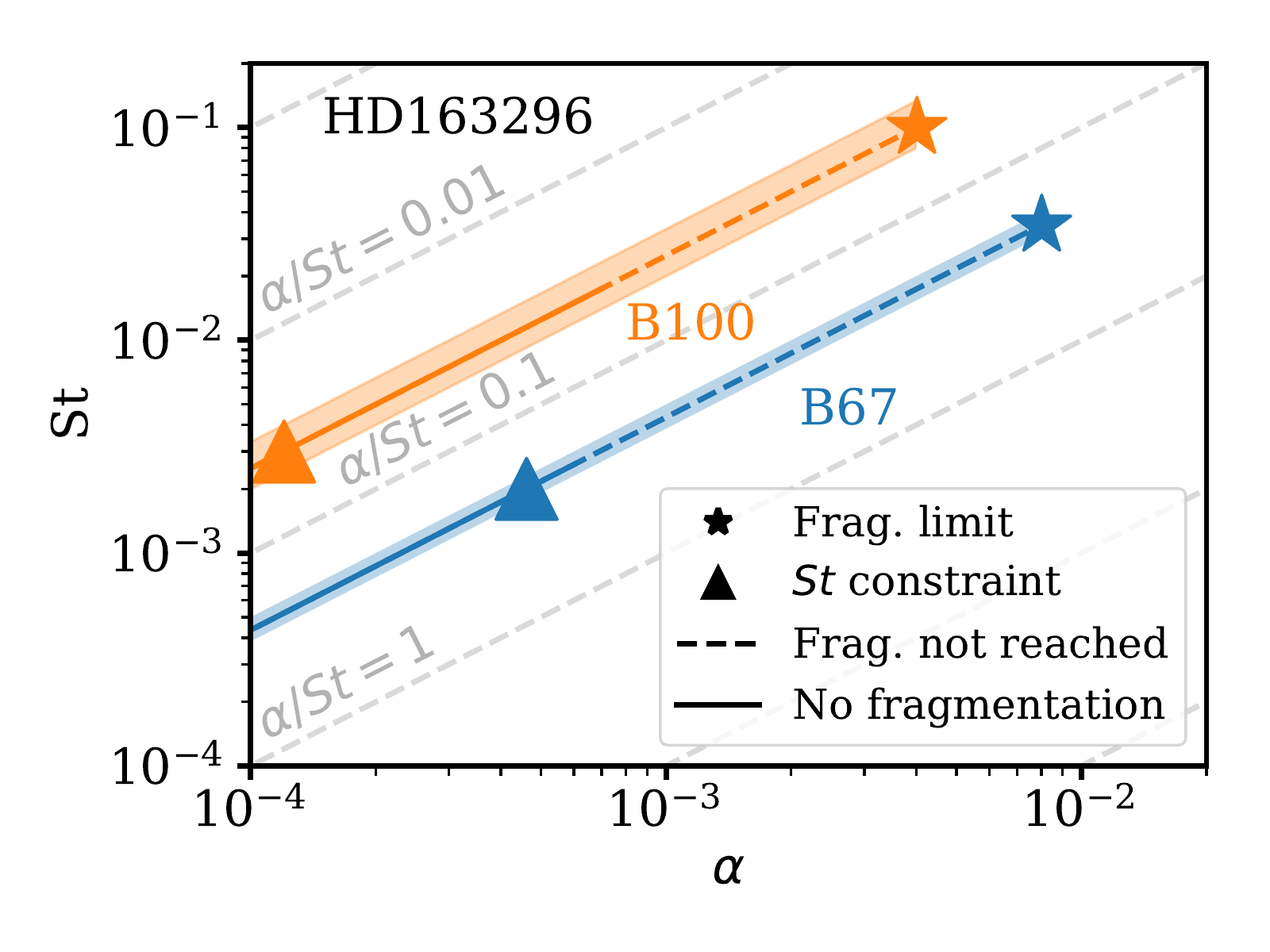}
\includegraphics[width=\columnwidth]{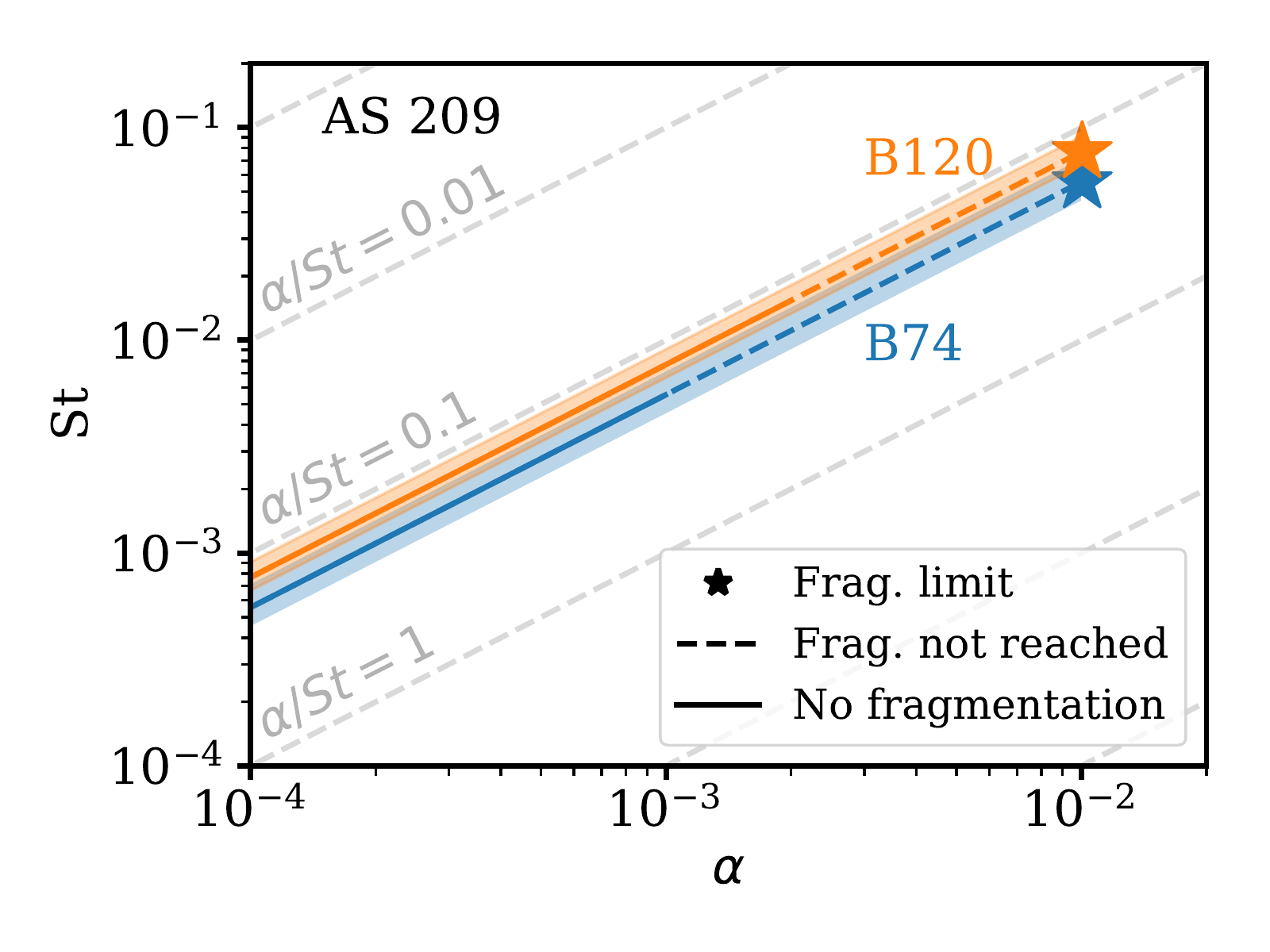}
\caption{Graphical visualisation of the constraints on $St$ and $\alpha$ derived in this paper. We do not plot B155 because the constraints overlap almost completely with the ones for B100. Our method yields a measurement of $\alpha/St$, i.e. a line in this plane (see grey dashed lines for reference lines of constant $\alpha/St$). We mark with the stars the fragmentation limit. The solid line marks the region where fragmentation due to turbulence never operates and therefore another process must limit the grain size. In the region marked with the dashed line, fragmentation due to turbulence is possible, but because of the lower $St$ another process is limiting the grain size more efficiently than fragmentation. For HD163296, we use the disc surface density profile of \citet{Booth2019} to set a constraint on $St$, assuming a grain size of 1 mm.}
\label{fig:stokes_alpha_plane}
\end{figure*}

We now quantify the efficiency of this stirring mechanism. We report in \autoref{tab:results} the resulting $\alpha/St$ values, deduced using \autoref{eq:alpha_St_main}, and we plot graphically the constraints in the $\alpha-St$ plane in \autoref{fig:stokes_alpha_plane}. In general, we find that the degree of coupling of the dust is relatively good, with an average $\alpha/St \sim 0.1$. We discuss the implications of these results in section \ref{sec:discussion_conclusion}.

Finally, we also measure the gas width $w_g$ using \autoref{eq:width_gas_main}. Although \citetalias{Dullemond2018} did not analyse the gas data to measure gas widths, they identified possible lower and upper limits based on different physical criteria. Our values fall inside this range except for B100, in which case we find a wider gap than their upper limit - this might be because they employed the \textit{full} width half-maximum to set a constraint on the possible width. We note that in \textit{all} cases the gas widths are larger than the dust ones, providing support to the idea that these structures are pressure traps.% On the other hand, 

%\rt{We should be careful here, what I've called $v_{\rm kep}$ in the table is actually the value from the double power-law fit which isn't truly Keplerian rotation. Should we use a different notation for this as to not confuse people? $v_{\rm background}$ or something less cumbersome? $v_{\rm bg}$?} \gr{I prefer to leave $v_{\rm kep}$ as a notation because it makes reading the table easier - but I've added a footnote in the previous section and a parenthesis here, to make sure that people are aware of what we are doing}

\subsection{Deriving an \texorpdfstring{$\alpha$}{α} value}

\begin{table}
\caption{Constraints on $\alpha$ and grain properties derived from our measurements of $\alpha/St$. For AS209, we do not use the surface density reported by \citet{Favre2019} to set a constraint on $St$ and $\alpha$ because the measurement is most likely affected by carbon depletion.}
\label{tab:dust}
%\centering
\begin{tabular}{lccccc} \hline\hline
 & (1) & (2) & (3) & (4) & (5) \\
Ring & $\alpha_\mathrm{frag}$ & $\alpha_\mathrm{min,frag}$  & $\Sigma_\mathrm{gas}$ & $St_\mathrm{\Sigma}$ & $\alpha_\mathrm{\Sigma}$ \\
 & & & ($\rm g \,cm^{-2}$) &   \\ \hline
 %$St_\mathrm{max}$ & $\Sigma_\mathrm{min}$
\multicolumn{6}{c}{\footnotesize HD~163296} \\ \hline
B67   &  $8 \times 10^{-3}$ & $6\times 10^{-4}$ &68& $2\times 10^{-3}$ & $6 \times 10^{-4}$\\
B100  &  $4 \times 10^{-3}$ & $7 \times 10^{-4}$ & 57 & $3\times 10^{-3}$ & $ 10^{-4}$\\
B155  &  $4 \times 10^{-3}$ & $10^{-3}$ & 53 & $3\times 10^{-3}$ & $10^{-4}$  \\
\hline
\multicolumn{6}{c}{\footnotesize AS~209} \\ \hline           
B74     & $10^{-2}$ & $10^{-3}$  & 0.3 & &  \\
B120    & $10^{-2}$ & $2 \times 10^{-3}$ & 0.3& & \\
\hline
%(${\rm m\,s^{-1}}$) & $533 \pm 27$ &  & $500 \pm 22$ & $488 \pm 18$ & $381 \pm 20$ & $359 \pm 14$
\end{tabular}
\par \textit{Notes} (1) Value of $\alpha$ if the grain size is set by fragmentation (2) Value of $\alpha$ below which fragmentation does not limit grain size (3) Gas surface density derived from observations (4) Stokes number with the previous value of the surface density, assuming a grain size of 1 mm. (5) $\alpha$ value obtained combining the previous constraint on $St$ and our measurement of $\alpha/St$.
\end{table}

Because the dust dynamics depends only on the ratio $\alpha/St$, so far we have been unable to measure individually the two parameters. To break the degeneracy between them, we need some information on $St$. As shown by \citet{Birnstiel2012}, in models of dust coagulation the dust grain size is limited by either fragmentation or radial drift. Since the rings are pressure maxima, the dust is not rapidly drifting in their vicinity; therefore, it is plausible to assume that the grain size should be set by fragmentation (e.g., see \citealt{Bae2018}). In this case, using Equation 3 of \citet{Birnstiel2012}:
\begin{equation}
    \alpha_\mathrm{frag}=\sqrt{\frac{1}{3} \frac{u_f^2}{c_s^2} \left(\frac{\alpha}{St}\right)_\mathrm{measured} },
    \label{eq:alpha_frag}
\end{equation}
where $u_f$ is the fragmentation velocity and $c_s$ the sound speed in the midplane \citetalias[estimated using the temperatures computed by][]{Dullemond2018}. We report in \autoref{tab:dust} the resulting $\alpha_\mathrm{frag}$ values when assuming a value of the fragmentation velocity of 10~${\rm m\,s^{-1}}$ (note the linear dependence on this parameter). We also added these values as the stars markers in \autoref{fig:stokes_alpha_plane}. Conventionally, $\alpha$ is assumed to lie in the approximate range $[10^{-4},10^{-2}]$; the values we find are towards the upper end of this range.

We cannot know if the grain size is indeed set by fragmentation, but we argue that $\alpha_\mathrm{frag}$ is an upper limit on the value of $\alpha$. This is because, if $\alpha$ was \textit{greater} than $\alpha_\mathrm{frag}$, fragmentation would limit the grain size to a $St$ not compatible with our measurement of $\alpha/St$. Instead, it is acceptable that $\alpha$ is \textit{lower} than $\alpha_\mathrm{frag}$ if we invoke the presence of another process (e.g., bouncing, or a residual level of radial drift) setting the grain size, and that this process limits the grain size to a value \textit{smaller} than what would be set by fragmentation. This is marked with the solid and dashed lines in \autoref{fig:stokes_alpha_plane} (see next paragraph for the difference between solid and dashed).%Because the grain size limited by fragmentation is inversely proportional to $\alpha$,  Conversely, 

Lastly, it should be noted that the relative velocity in grain collisions increases with $St$ but only until $St=1$ \citep{OrmelCuzzi2007}; increasing $St$ further \textit{decreases} the relative velocity. This has two important practical consequences. Firstly, \autoref{eq:alpha_frag} assumes that the relative velocity always increases with $St$ and therefore is only valid for the case $St<1$; we have verified \textit{a posteriori} that in all cases we obtain an acceptable solution, i.e. with $St<1$. % While this means that the $\alpha_\mathrm{frag}$ values we report are acceptable solutions, we remark that they are not the only possible ones: 
Secondly, if $\alpha$ is sufficiently low, even for $St=1$ the relative velocity is lower than the fragmentation velocity and therefore the fragmentation limit never applies. We define $\alpha_\mathrm{min,frag}$ as this critical value of $\alpha$; its value is $\alpha_\mathrm{min,frag} = 2/3 \, u_f^2/c_s^2$ (note that in this case the dependence on $u_f$ is quadratic); as previously, we compute these values assuming a value of the fragmentation velocity of 10~${\rm m\,s^{-1}}$ and report them in \autoref{tab:dust}. These values are the separation between the solid and dashed line in \autoref{fig:stokes_alpha_plane}. The significance of the solid region is that, even without our measurements of $\alpha/St$, in this region we \textit{must} invoke another process limiting grain growth, or growth would proceed unimpeded. %In principle, another process limiting grain size can be invoked also in the intermediate region $\alpha_\mathrm{max,frag} < \alpha < \alpha_\mathrm{frag}$, and in this case the process must be more efficient than fragmentation.

To summarise, there are two possible scenarios; in the first, the grain size is set by fragmentation and $\alpha$ takes the value we report as $\alpha_\mathrm{frag}$. It is interesting to note that for HD163296 these values are incompatible with the upper limit reported by \citet{Flaherty2017} of $3 \times 10^{-3}$, especially for B67, while for B100 and B155 a slight reduction in fragmentation velocity could still make the two measurements compatible. In the second scenario, another process is setting the grain size and $\alpha$ can take any value \textit{smaller} than $\alpha_\mathrm{frag}$. In this case, depending on the value of $\alpha$, we can also further argue that if $\alpha_\mathrm{min,frag} < \alpha < \alpha_\mathrm{frag}$ this process must be more efficient than fragmentation.% Conversely, no constraints on the grain-limiting process can be set if $\alpha < \alpha_\mathrm{max,frag}$.

\subsection{Which Stokes numbers are compatible with grain growth?}
\label{sec:disc_mass}

Because $St$ is linked to the grain size, it is worth asking what values are compatible with the well known results of dust grain growth in proto-planetary discs (see \citealt{Testi2014} for a review); in turn, this sets a constraint on $\alpha$ given the measurements of $\alpha/St$ that we present in this paper. The Stokes number in the Epstein regime can be expressed as:
\begin{equation}
St = 1.5 \times 10^{-3} \left( \frac{a}{1 \ \mathrm{mm}}\right) \left( \frac{\Sigma}{100 \ \mathrm{g \ cm^{-2}}} \right)^{-1},
\end{equation}
where $a$ is the grain size and we have assumed a dust bulk density of $1 \ \mathrm{g \ cm^{-3}}$. To put some constraint on $St$, we thus need measurements of the gas surface density.

%To study this, we have added a column in \autoref{tab:dust}, indicating the \textit{maximum} possible value of $St$ if $\alpha=\alpha_\mathrm{min,frag}$. The Stokes number depends on grain size and gas surface density; for reference purposes, we report in \autoref{tab:results} the \textit{minimum} value of the surface density $\Sigma_\mathrm{minimum}$ (computed assuming a dust bulk density of 1 $\rm g \, cm^{-3}$) compatible with 1 $\rm mm$ grains. 

For HD163296, such a measurement is provided by the recent detection of $^{13}$C$^{17}$O \citep{Booth2019}. Given the non-detection of this disc in the HD 1-0 transition \citep{Kama2020}, the gas surface density of this disc is reasonably well constrained, since increasing it would make it incompatible with the non-detection of HD and gravitationally unstable (in contrast with the lack of observed spiral arms), while lowering it would make it incompatible with the detection of $^{13}$C$^{17}$O. We list in \autoref{tab:dust} the surface density $\Sigma_\mathrm{gas}$ at each ring location from the disc model of \citet{Booth2019}; we then use this surface density to compute the resulting Stokes number $St_\mathrm{\Sigma}$. These values are plotted as the triangles in \autoref{fig:stokes_alpha_plane}. The resulting $\alpha$ (which we list as $\alpha_\mathrm{Sigma}$ in the table), once combined with our measurements of $\alpha/St$, seem to exclude the possibility that $\alpha$ is high and that the grain size is set by fragmentation. In order to make fragmentation the process setting grain size, we would need a grain size larger by a factor $\gtrsim$ 10 to increase $\alpha_\mathrm{Sigma}$, or a fragmentation velocity smaller by a similar amount to decrease $\alpha_\mathrm{frag}$ (or a combination of both).

For AS209 instead, \citet{Favre2019} reports significantly lower values of the surface density from CO isotopologues observations. At face value, this would point towards the need for much larger $St$ ($\sim$ 0.5), that are not compatible with our constraints since they would imply a value of $\alpha$ greater than $\alpha_\mathrm{frag}$. Additionally, this result would be at odds with attempts at modelling the dust structure in AS~209 as due to disc-planet interaction, that consistently highlighted the need for low viscosity values in this particular disc \citep{Fedele2018,Zhang2018}. However, it is well known that due to carbon depletion CO-derived disc masses are generally underestimated in T Tauri stars \citep[e.g.,][]{Miotello2017} and therefore it is likely that the true disc mass is significantly higher than the estimate of \citet{Favre2019}. For this reason, we do not plot these constraints in \autoref{fig:stokes_alpha_plane}, nor indicate them in \autoref{tab:dust}.

A caveat of this analysis is that we have simply assumed that the grains are 1 $\rm mm$. A better estimate would be needed, but we note that, even if ALMA has now been in operation for a few years, very few discs have been studied with sufficient spatial resolution at multiple wavelengths to study the grain properties in the rings, and therefore we have limited information on the grain size. For example, for HD163296 the spatial resolution of \citet{Guidi2016} was not enough to resolve the rings; \citet{Dent2019} was not able to place constraints on the grain size due to the degeneracy between grain growth and optical depth, and the limited difference in wavelength between band 6 and band 7. While polarisation could in principle be an alternative way of placing constraints on the grain size, the analysis of \citet{Ohashi2019} shows that the potentially high optical depth of the rings in the sub-mm makes the grain size unconstrained, highlighting the need for data at longer wavelengths. Future high-resolution studies will provide constraints on the grain size, in this way further breaking the degeneracy between $\alpha$ and $St$.

\subsection{Caveats}
\label{sec:caveats}

%Our analysis is based on a first order expansion of the logarithmic derivative of the pressure profile. This is equivalent to assuming that the gas pressure profile is described by a Gaussian (see section \ref{sec:width_gas}). Such an assumption is certainly a good \textit{local} description, but it certainly cannot be a good \textit{global} description given that we observe multiple dust rings and the deviation from Keplerian is no longer a straight line sufficiently far from the pressure maximum.

As we highlighted when describing \autoref{fig:vrot_observations}, the rotation curve we derive from the data is broadly consistent with the dust continuum structure. However, we wish to discuss in this section two effects that are in partial tension with the dust continuum. The first one has been already introduced in section \ref{sec:derived_width}, namely that for B100 in HD163296 and B120 for AS~209 the continuum peak is not located at the center of the radial range over which $\delta v_{\phi}$ decreases. The problem is particularly severe for AS~209, in which case $\delta v_{\phi}$ starts decreasing only at the location of the peak in the continuum. For AS~209, this effect has already been noted by \citet{Teague2018AS209}. We note that, because we do not know the true value of the Keplerian velocity, there is some uncertainty in the exact location of the pressure maximum (i.e., a constant vertical offset in $\delta v_{\phi}$ would shift radially the location where the pressure gradient crosses zero, see \citealt{Keppler2019} for an example). While this could be enough to explain the inconsistency for B100 in HD163296, it does not appear to be the case for B120 for AS~209, since a vertical offset would not change the fact that $\delta v_{\phi}$ increases (i.e., has a positive derivative) at radii smaller than inside the continuum peak. 

There are two reasons that could explain this discrepancy. The first one is what we discuss in appendix \ref{sec:vertical_structure}, namely the effect of a local variation in the height of the emission surface. \autoref{fig:perturb} shows an example where the deviation from Keplerian starts decreasing only outside the location of the pressure maximum. The second reason is the possibility that the gap structure is intrinsically not symmetrical. Within the framework of this paper, we cannot account for an asymmetry because, to first order in $r-r_0$, the gap structure is symmetrical by construction, but this is a possibility we plan to investigate in future papers. The issue is of interest because hydro-dynamical models of disc-planet interaction tend to predict a steeper pressure profile inside the pressure maximum than outside. It is also suggestive that observational studies of transition discs show \citep{Pinilla2018} a similar difference in the dust distribution on the two sides of the pressure maximum. Therefore, both effects go in the same direction. With the current data, it is not currently possible to disentangle between them.

The second caveat we wish to discuss concerns the magnitude of the deviation from Keplerian. We can hypothesise that at sufficient distance from the pressure bump the pressure profile goes back to some smooth, negative slope and therefore the rotation curve is sub-Keplerian. As already discussed in this paper, we cannot measure this unperturbed slope because of the uncertainties in the mass of the star, as well as the height of the emission surface. However, we can write that in the unperturbed region the deviation from Keplerian should be of order $\delta v_{\phi}/v_K = 1/2 (c_s/v_K)^2 \gamma \simeq (c_s/v_K)^2$ (see \autoref{eq:vphi_order}), where we have called $\gamma$ the logarithmic slope of the unperturbed surface density profile, and in the last passage we have ignored factors of order unity. For the pressure bump to be a pressure maximum, the change in $v_{\phi}$ induced by the bump needs to be high enough for the rotation curve to transition from sub- to super-Keplerian rotation. Given a radial range $\Delta r$ of the variation and a slope $m=\Delta \delta v_{\phi} / \Delta r$, this means that the total variation $m \Delta r$ induced by the pressure bump needs to be larger (in absolute value) than the unperturbed value $(c_s/v_K)^2$ (note that, because of the analysis of appendix \ref{sec:vertical_structure}, it does not matter whether we perform this comparison in the midplane or at the emission surface). We list in \autoref{tab:results} the $\Delta r$ we employ and the ratio between the total variation and $(c_s/v_K)^2$. It can be seen that, whereas in AS~209 the total variation is comfortably higher than what is needed to produce pressure maxima, mostly because of the larger slopes measured from the data, for HD163296 the total variation is barely larger than the constraint; for B155, the slope we measure is not large enough to produce a pressure maximum. Even for the other two rings, this does not leave much free room to have a pressure maximum. This could be because the pressure bumps in this disc are indeed only shallow maxima, or because $\gamma$ is small (i.e., the unperturbed surface density is very shallow), or it could be because of additional physics we are missing in our analysis.

It should be remarked that the datasets we analysed were not designed to study gas kinematics; for example they have a quite limited spectral resolution (0.35 km/s native resolution, with the actual resolution roughly two times worse due to Hanning smoothing). Ultimately, separate datasets explicitly designed to study kinematics are needed to re-assess in future works the caveats we describe here.

\section{Discussion and conclusions}
\label{sec:discussion_conclusion}

In this paper we presented a unique approach to analyse the disc kinematics and, in comparison with the sub-mm continuum emission, measure the $\alpha/St$ ratio at the ring centres, in this way providing constraints on the level of turbulence and the dust-gas coupling. Moreover, our method also measures the width of gas pressure bumps. Our results confirm that the structures in the gas are larger than in the dust and that $\alpha/St<1$, thereby providing evidence that the rings now ubiquitously imaged are dust traps, at least for the two discs studied here.

At the same time, our results also imply a relatively large value of $\alpha/St$, with a typical value of 0.1. Our constraints are illustrated in \autoref{fig:stokes_alpha_plane} and they imply that we can reject a scenario in which the disc is characterised by low turbulence (e.g., $\alpha=10^{-4}$) and the grains have large Stokes numbers (e.g., $St=0.1$). On the contrary, our results imply that if the grains have large Stokes numbers then the disc must also be very turbulent (at least in the radial direction), for example in the case limited by fragmentation (see $\alpha_\mathrm{frag}$ values in \autoref{tab:results}). This case also constitutes an upper limit on the value of $\alpha$. On the other hand, such high values of the turbulence appear to be in tension with the lack of a direct detection \citep{Flaherty2017}, at least for the case of HD~163296. We note that the analysis carried by \citet{Flaherty2017} assumed homogeneous, isotropic turbulence, and it is possible that this discrepancy may be solved by relaxing this assumption. The other possibility is instead that the discrepancy points to a different physical regime, namely that the grain size in these discs is not set by fragmentation ($\alpha$ smaller than $\alpha_\mathrm{frag}$ in \autoref{tab:dust}). This possibility is more in line with recent theoretical work proposing that accretion is mostly driven by winds launched by the magnetic field. This option is compatible with our data and, as we discuss in section \ref{sec:disc_mass}, also with recent measurements of the disc mass \citep{Booth2019} for HD163296. For AS 209 instead, our results are in tension with the low disc mass inferred from C$^{18}$O observations \citep{Favre2019}, although it is likely that carbon depletion is severely affecting those measurements.

Lastly, it should be noted that future high-resolution gas observations of optically thin lines (which for the two discs we analysed will be conducted by the approved Large Programme MAPS) will test our measurements of gas widths and will provide an independent constraint. %As we have discussed in section \ref{sec:surface_density_assumption}, the gas gaps could be narrower than what we have derived here if they have a small depth. 
We remark that the method we propose here is cheaper in terms of observing time since it requires brighter, optically thick lines. A validation of our method would then allow to apply it to a larger disc sample.

%\begin{itemize}
    %\item We confirm that structures in the gas are larger than in the dust. This is a confirmation of dust trapping
    %\item However, difference is not big, implying low $St/\alpha$: either discs are highly turbulent, or the grains have small sizes. 
%    \item Our gas widths are already as large as they can be; therefore these constraints come from the large widths of the features in the dust.
    %\item Future high-resolution campaigns at longer wavelengths (e.g., 3mm) will inform us about the grain size, partially breaking the degeneracy between grain size/Stokes number and $\alpha$
    %\item Discrepancy with flaherty
%\end{itemize}

\section*{Acknowledgements}
This work is part of the research programme VENI with project number 016.Veni.192.233, which is (partly) financed by the Dutch Research Council (NWO). RB and CJC acknowledge support from the STFC consolidated grant ST/S000623/1. This work has also been supported by the European Union’s Horizon 2020 research and innovation programme under the Marie Sklodowska-Curie grant agreement No 823823 (DUSTBUSTERS).

%%%%%%%%%%%%%%%%%%%%%%%%%%%%%%%%%%%%%%%%%%%%%%%%%%

%%%%%%%%%%%%%%%%%%%% REFERENCES %%%%%%%%%%%%%%%%%%

% The best way to enter references is to use BibTeX:

\bibliographystyle{mnras}
\bibliography{diffusivity} % if your bibtex file is called example.bib

\begin{thebibliography}{}
\makeatletter
\relax
\def\mn@urlcharsother{\let\do\@makeother \do\$\do\&\do\#\do\^\do\_\do\%\do\~}
\def\mn@doi{\begingroup\mn@urlcharsother \@ifnextchar [ {\mn@doi@}
  {\mn@doi@[]}}
\def\mn@doi@[#1]#2{\def\@tempa{#1}\ifx\@tempa\@empty \href
  {http://dx.doi.org/#2} {doi:#2}\else \href {http://dx.doi.org/#2} {#1}\fi
  \endgroup}
\def\mn@eprint#1#2{\mn@eprint@#1:#2::\@nil}
\def\mn@eprint@arXiv#1{\href {http://arxiv.org/abs/#1} {{\tt arXiv:#1}}}
\def\mn@eprint@dblp#1{\href {http://dblp.uni-trier.de/rec/bibtex/#1.xml}
  {dblp:#1}}
\def\mn@eprint@#1:#2:#3:#4\@nil{\def\@tempa {#1}\def\@tempb {#2}\def\@tempc
  {#3}\ifx \@tempc \@empty \let \@tempc \@tempb \let \@tempb \@tempa \fi \ifx
  \@tempb \@empty \def\@tempb {arXiv}\fi \@ifundefined
  {mn@eprint@\@tempb}{\@tempb:\@tempc}{\expandafter \expandafter \csname
  mn@eprint@\@tempb\endcsname \expandafter{\@tempc}}}

\bibitem[\protect\citeauthoryear{{ALMA Partnership} et~al.,}{{ALMA Partnership}
  et~al.}{2015}]{2015ApJ...808L...3A}
{ALMA Partnership} et~al., 2015, \mn@doi [\apjl] {10.1088/2041-8205/808/1/L3},
  \href {https://ui.adsabs.harvard.edu/abs/2015ApJ...808L...3A} {808, L3}

\bibitem[\protect\citeauthoryear{{Andrews} et~al.,}{{Andrews}
  et~al.}{2018}]{Andrews2018}
{Andrews} S.~M.,  et~al., 2018, \mn@doi [\apjl] {10.3847/2041-8213/aaf741},
  \href {https://ui.adsabs.harvard.edu/abs/2018ApJ...869L..41A} {869, L41}

\bibitem[\protect\citeauthoryear{{Bae}, {Pinilla}  \& {Birnstiel}}{{Bae}
  et~al.}{2018}]{Bae2018}
{Bae} J.,  {Pinilla} P.,   {Birnstiel} T.,  2018, \mn@doi [\apjl]
  {10.3847/2041-8213/aadd51}, \href
  {https://ui.adsabs.harvard.edu/abs/2018ApJ...864L..26B} {864, L26}

\bibitem[\protect\citeauthoryear{{Birnstiel}, {Klahr}  \&
  {Ercolano}}{{Birnstiel} et~al.}{2012}]{Birnstiel2012}
{Birnstiel} T.,  {Klahr} H.,   {Ercolano} B.,  2012, \mn@doi [\aap]
  {10.1051/0004-6361/201118136}, \href
  {https://ui.adsabs.harvard.edu/abs/2012A&A...539A.148B} {539, A148}

\bibitem[\protect\citeauthoryear{{Bodenheimer}, {D'Angelo}, {Lissauer},
  {Fortney}  \& {Saumon}}{{Bodenheimer} et~al.}{2013}]{Bodenheimer2013}
{Bodenheimer} P.,  {D'Angelo} G.,  {Lissauer} J.~J.,  {Fortney} J.~J.,
  {Saumon} D.,  2013, \mn@doi [\apj] {10.1088/0004-637X/770/2/120}, \href
  {https://ui.adsabs.harvard.edu/abs/2013ApJ...770..120B} {770, 120}

\bibitem[\protect\citeauthoryear{{Booth}, {Walsh}, {Ilee}, {Notsu}, {Qi},
  {Nomura}  \& {Akiyama}}{{Booth} et~al.}{2019}]{Booth2019}
{Booth} A.~S.,  {Walsh} C.,  {Ilee} J.~D.,  {Notsu} S.,  {Qi} C.,  {Nomura} H.,
    {Akiyama} E.,  2019, \mn@doi [\apjl] {10.3847/2041-8213/ab3645}, \href
  {https://ui.adsabs.harvard.edu/abs/2019ApJ...882L..31B} {882, L31}

\bibitem[\protect\citeauthoryear{{Brauer}, {Dullemond}, {Johansen}, {Henning},
  {Klahr}  \& {Natta}}{{Brauer} et~al.}{2007}]{Brauer2007}
{Brauer} F.,  {Dullemond} C.~P.,  {Johansen} A.,  {Henning} T.,  {Klahr} H.,
  {Natta} A.,  2007, \mn@doi [\aap] {10.1051/0004-6361:20066865}, \href
  {https://ui.adsabs.harvard.edu/abs/2007A&A...469.1169B} {469, 1169}

\bibitem[\protect\citeauthoryear{{Clarke} et~al.,}{{Clarke}
  et~al.}{2018}]{Clarke2018}
{Clarke} C.~J.,  et~al., 2018, \mn@doi [\apjl] {10.3847/2041-8213/aae36b},
  \href {https://ui.adsabs.harvard.edu/abs/2018ApJ...866L...6C} {866, L6}

\bibitem[\protect\citeauthoryear{{Dartois}, {Dutrey}  \&
  {Guilloteau}}{{Dartois} et~al.}{2003}]{Dartois2003}
{Dartois} E.,  {Dutrey} A.,   {Guilloteau} S.,  2003, \mn@doi [\aap]
  {10.1051/0004-6361:20021638}, \href
  {https://ui.adsabs.harvard.edu/abs/2003A&A...399..773D} {399, 773}

\bibitem[\protect\citeauthoryear{{Dent}, {Pinte}, {Cortes}, {M{\'e}nard},
  {Hales}, {Fomalont}  \& {de Gregorio-Monsalvo}}{{Dent}
  et~al.}{2019}]{Dent2019}
{Dent} W.~R.~F.,  {Pinte} C.,  {Cortes} P.~C.,  {M{\'e}nard} F.,  {Hales} A.,
  {Fomalont} E.,   {de Gregorio-Monsalvo} I.,  2019, \mn@doi [\mnras]
  {10.1093/mnrasl/sly181}, \href
  {https://ui.adsabs.harvard.edu/abs/2019MNRAS.482L..29D} {482, L29}

\bibitem[\protect\citeauthoryear{{Dipierro} et~al.,}{{Dipierro}
  et~al.}{2018}]{Dipierro2018}
{Dipierro} G.,  et~al., 2018, \mn@doi [\mnras] {10.1093/mnras/sty181}, \href
  {https://ui.adsabs.harvard.edu/abs/2018MNRAS.475.5296D} {475, 5296}

\bibitem[\protect\citeauthoryear{{Dullemond} et~al.,}{{Dullemond}
  et~al.}{2018}]{Dullemond2018}
{Dullemond} C.~P.,  et~al., 2018, \mn@doi [\apj] {10.3847/2041-8213/aaf742},
  \href {https://ui.adsabs.harvard.edu/abs/2018ApJ...869L..46D} {869, L46}

\bibitem[\protect\citeauthoryear{{Dullemond}, {Isella}, {Andrews}, {Skobleva}
  \& {Dzyurkevich}}{{Dullemond} et~al.}{2020}]{Dullemond2019}
{Dullemond} C.,  {Isella} A.,  {Andrews} S.,  {Skobleva} I.,   {Dzyurkevich}
  N.,  2020, \aap, \href
  {https://ui.adsabs.harvard.edu/abs/2019arXiv191112434D} {633, A137}

\bibitem[\protect\citeauthoryear{{Facchini} et~al.,}{{Facchini}
  et~al.}{2019}]{Facchini2019}
{Facchini} S.,  et~al., 2019, \mn@doi [\aap] {10.1051/0004-6361/201935496},
  \href {https://ui.adsabs.harvard.edu/abs/2019A&A...626L...2F} {626, L2}

\bibitem[\protect\citeauthoryear{{Favre} et~al.,}{{Favre}
  et~al.}{2019}]{Favre2019}
{Favre} C.,  et~al., 2019, \mn@doi [\apj] {10.3847/1538-4357/aaf80c}, \href
  {https://ui.adsabs.harvard.edu/abs/2019ApJ...871..107F} {871, 107}

\bibitem[\protect\citeauthoryear{{Fedele} et~al.,}{{Fedele}
  et~al.}{2017}]{Fedele2017}
{Fedele} D.,  et~al., 2017, \mn@doi [\aap] {10.1051/0004-6361/201629860}, \href
  {https://ui.adsabs.harvard.edu/abs/2017A&A...600A..72F} {600, A72}

\bibitem[\protect\citeauthoryear{{Fedele} et~al.,}{{Fedele}
  et~al.}{2018}]{Fedele2018}
{Fedele} D.,  et~al., 2018, \mn@doi [\aap] {10.1051/0004-6361/201731978}, \href
  {https://ui.adsabs.harvard.edu/abs/2018A&A...610A..24F} {610, A24}

\bibitem[\protect\citeauthoryear{{Flaherty} et~al.,}{{Flaherty}
  et~al.}{2017}]{Flaherty2017}
{Flaherty} K.~M.,  et~al., 2017, \mn@doi [\apj] {10.3847/1538-4357/aa79f9},
  \href {https://ui.adsabs.harvard.edu/abs/2017ApJ...843..150F} {843, 150}

\bibitem[\protect\citeauthoryear{{Flock}, {Fromang}, {Gonz{\'a}lez}  \&
  {Commer{\c{c}}on}}{{Flock} et~al.}{2013}]{Flock2013}
{Flock} M.,  {Fromang} S.,  {Gonz{\'a}lez} M.,   {Commer{\c{c}}on} B.,  2013,
  \mn@doi [\aap] {10.1051/0004-6361/201322451}, \href
  {https://ui.adsabs.harvard.edu/abs/2013A&A...560A..43F} {560, A43}

\bibitem[\protect\citeauthoryear{{Greaves} \& {Rice}}{{Greaves} \&
  {Rice}}{2010}]{Greaves2010}
{Greaves} J.~S.,  {Rice} W.~K.~M.,  2010, \mn@doi [\mnras]
  {10.1111/j.1365-2966.2010.17043.x}, \href
  {https://ui.adsabs.harvard.edu/abs/2010MNRAS.407.1981G} {407, 1981}

\bibitem[\protect\citeauthoryear{{Guidi} et~al.,}{{Guidi}
  et~al.}{2016}]{Guidi2016}
{Guidi} G.,  et~al., 2016, \mn@doi [\aap] {10.1051/0004-6361/201527516}, \href
  {https://ui.adsabs.harvard.edu/abs/2016A&A...588A.112G} {588, A112}

\bibitem[\protect\citeauthoryear{{Guzm{\'a}n} et~al.,}{{Guzm{\'a}n}
  et~al.}{2018}]{Guzman2018}
{Guzm{\'a}n} V.~V.,  et~al., 2018, \mn@doi [\apjl] {10.3847/2041-8213/aaedae},
  \href {https://ui.adsabs.harvard.edu/abs/2018ApJ...869L..48G} {869, L48}

\bibitem[\protect\citeauthoryear{{Huang} et~al.,}{{Huang}
  et~al.}{2018}]{Huang2018}
{Huang} J.,  et~al., 2018, \mn@doi [\apjl] {10.3847/2041-8213/aaf740}, \href
  {https://ui.adsabs.harvard.edu/abs/2018ApJ...869L..42H} {869, L42}

\bibitem[\protect\citeauthoryear{{Isella} et~al.,}{{Isella}
  et~al.}{2018}]{Isella2018}
{Isella} A.,  et~al., 2018, \mn@doi [\apjl] {10.3847/2041-8213/aaf747}, \href
  {https://ui.adsabs.harvard.edu/abs/2018ApJ...869L..49I} {869, L49}

\bibitem[\protect\citeauthoryear{{Kama} et~al.,}{{Kama}
  et~al.}{2019}]{Kama2020}
{Kama} M.,  et~al., 2019, arXiv e-prints, \href
  {https://ui.adsabs.harvard.edu/abs/2019arXiv191211883K} {p. arXiv:1912.11883}

\bibitem[\protect\citeauthoryear{{Keppler} et~al.,}{{Keppler}
  et~al.}{2019}]{Keppler2019}
{Keppler} M.,  et~al., 2019, \mn@doi [\aap] {10.1051/0004-6361/201935034},
  \href {https://ui.adsabs.harvard.edu/abs/2019A&A...625A.118K} {625, A118}

\bibitem[\protect\citeauthoryear{{Kley} \& {Nelson}}{{Kley} \&
  {Nelson}}{2012}]{2012ARA&A..50..211K}
{Kley} W.,  {Nelson} R.~P.,  2012, \mn@doi [\araa]
  {10.1146/annurev-astro-081811-125523}, \href
  {https://ui.adsabs.harvard.edu/abs/2012ARA&A..50..211K} {50, 211}

\bibitem[\protect\citeauthoryear{{Lodato} et~al.,}{{Lodato}
  et~al.}{2019}]{Lodato2019}
{Lodato} G.,  et~al., 2019, \mn@doi [\mnras] {10.1093/mnras/stz913}, \href
  {https://ui.adsabs.harvard.edu/abs/2019MNRAS.486..453L} {486, 453}

\bibitem[\protect\citeauthoryear{{Long} et~al.,}{{Long}
  et~al.}{2018}]{Long2018}
{Long} F.,  et~al., 2018, \mn@doi [\apj] {10.3847/1538-4357/aae8e1}, \href
  {https://ui.adsabs.harvard.edu/abs/2018ApJ...869...17L} {869, 17}

\bibitem[\protect\citeauthoryear{{Long} et~al.,}{{Long}
  et~al.}{2019}]{Long2019}
{Long} F.,  et~al., 2019, arXiv e-prints, \href
  {https://ui.adsabs.harvard.edu/abs/2019arXiv190610809L} {p. arXiv:1906.10809}

\bibitem[\protect\citeauthoryear{{Lynden-Bell} \& {Pringle}}{{Lynden-Bell} \&
  {Pringle}}{1974}]{1974MNRAS.168..603L}
{Lynden-Bell} D.,  {Pringle} J.~E.,  1974, \mn@doi [\mnras]
  {10.1093/mnras/168.3.603}, \href
  {https://ui.adsabs.harvard.edu/abs/1974MNRAS.168..603L} {168, 603}

\bibitem[\protect\citeauthoryear{{Manara}, {Morbidelli}  \& {Guillot}}{{Manara}
  et~al.}{2018}]{Manara2018}
{Manara} C.~F.,  {Morbidelli} A.,   {Guillot} T.,  2018, \mn@doi [\aap]
  {10.1051/0004-6361/201834076}, \href
  {https://ui.adsabs.harvard.edu/abs/2018A&A...618L...3M} {618, L3}

\bibitem[\protect\citeauthoryear{{Miotello} et~al.,}{{Miotello}
  et~al.}{2017}]{Miotello2017}
{Miotello} A.,  et~al., 2017, \mn@doi [\aap] {10.1051/0004-6361/201629556},
  \href {https://ui.adsabs.harvard.edu/abs/2017A&A...599A.113M} {599, A113}

\bibitem[\protect\citeauthoryear{{Ohashi} \& {Kataoka}}{{Ohashi} \&
  {Kataoka}}{2019}]{Ohashi2019}
{Ohashi} S.,  {Kataoka} A.,  2019, \mn@doi [\apj] {10.3847/1538-4357/ab5107},
  \href {https://ui.adsabs.harvard.edu/abs/2019ApJ...886..103O} {886, 103}

\bibitem[\protect\citeauthoryear{{Ormel} \& {Cuzzi}}{{Ormel} \&
  {Cuzzi}}{2007}]{OrmelCuzzi2007}
{Ormel} C.~W.,  {Cuzzi} J.~N.,  2007, \mn@doi [\aap]
  {10.1051/0004-6361:20066899}, \href
  {https://ui.adsabs.harvard.edu/abs/2007A&A...466..413O} {466, 413}

\bibitem[\protect\citeauthoryear{{P{\'e}rez} et~al.,}{{P{\'e}rez}
  et~al.}{2016}]{Perez2016}
{P{\'e}rez} L.~M.,  et~al., 2016, \mn@doi [Science] {10.1126/science.aaf8296},
  \href {https://ui.adsabs.harvard.edu/abs/2016Sci...353.1519P} {353, 1519}

\bibitem[\protect\citeauthoryear{{Pinilla} et~al.,}{{Pinilla}
  et~al.}{2018}]{Pinilla2018}
{Pinilla} P.,  et~al., 2018, \mn@doi [\apj] {10.3847/1538-4357/aabf94}, \href
  {https://ui.adsabs.harvard.edu/abs/2018ApJ...859...32P} {859, 32}

\bibitem[\protect\citeauthoryear{{Pinte}, {Dent}, {M{\'e}nard}, {Hales},
  {Hill}, {Cortes}  \& {de Gregorio-Monsalvo}}{{Pinte}
  et~al.}{2016}]{Pinte2016}
{Pinte} C.,  {Dent} W.~R.~F.,  {M{\'e}nard} F.,  {Hales} A.,  {Hill} T.,
  {Cortes} P.,   {de Gregorio-Monsalvo} I.,  2016, \mn@doi [\apj]
  {10.3847/0004-637X/816/1/25}, \href
  {https://ui.adsabs.harvard.edu/abs/2016ApJ...816...25P} {816, 25}

\bibitem[\protect\citeauthoryear{{Pinte} et~al.,}{{Pinte}
  et~al.}{2018a}]{Pinte2018}
{Pinte} C.,  et~al., 2018a, \mn@doi [\aap] {10.1051/0004-6361/201731377}, \href
  {https://ui.adsabs.harvard.edu/abs/2018A&A...609A..47P} {609, A47}

\bibitem[\protect\citeauthoryear{{Pinte} et~al.,}{{Pinte}
  et~al.}{2018b}]{Pinte2018IMLup}
{Pinte} C.,  et~al., 2018b, \mn@doi [\aap] {10.1051/0004-6361/201731377}, \href
  {https://ui.adsabs.harvard.edu/abs/2018A&A...609A..47P} {609, A47}

\bibitem[\protect\citeauthoryear{{Rosotti}, {Juhasz}, {Booth}  \&
  {Clarke}}{{Rosotti} et~al.}{2016}]{Rosotti2016}
{Rosotti} G.~P.,  {Juhasz} A.,  {Booth} R.~A.,   {Clarke} C.~J.,  2016, \mn@doi
  [\mnras] {10.1093/mnras/stw691}, \href
  {https://ui.adsabs.harvard.edu/abs/2016MNRAS.459.2790R} {459, 2790}

\bibitem[\protect\citeauthoryear{{Rosotti}, {Booth}, {Tazzari}, {Clarke},
  {Lodato}  \& {Testi}}{{Rosotti} et~al.}{2019}]{Rosotti2019}
{Rosotti} G.~P.,  {Booth} R.~A.,  {Tazzari} M.,  {Clarke} C.,  {Lodato} G.,
  {Testi} L.,  2019, \mn@doi [\mnras] {10.1093/mnrasl/slz064}, \href
  {https://ui.adsabs.harvard.edu/abs/2019MNRAS.486L..63R} {486, L63}

\bibitem[\protect\citeauthoryear{{Semenov} \& {Wiebe}}{{Semenov} \&
  {Wiebe}}{2011}]{SemenovWiebe2011}
{Semenov} D.,  {Wiebe} D.,  2011, \mn@doi [\apjs] {10.1088/0067-0049/196/2/25},
  \href {https://ui.adsabs.harvard.edu/abs/2011ApJS..196...25S} {196, 25}

\bibitem[\protect\citeauthoryear{{Shakura} \& {Sunyaev}}{{Shakura} \&
  {Sunyaev}}{1973}]{1973A&A....24..337S}
{Shakura} N.~I.,  {Sunyaev} R.~A.,  1973, \aap, \href
  {https://ui.adsabs.harvard.edu/abs/1973A&A....24..337S} {500, 33}

\bibitem[\protect\citeauthoryear{{Takeuchi} \& {Lin}}{{Takeuchi} \&
  {Lin}}{2002}]{TakeuchiLin2002}
{Takeuchi} T.,  {Lin} D.~N.~C.,  2002, \mn@doi [\apj] {10.1086/344437}, \href
  {https://ui.adsabs.harvard.edu/abs/2002ApJ...581.1344T} {581, 1344}

\bibitem[\protect\citeauthoryear{{Takeuchi} \& {Lin}}{{Takeuchi} \&
  {Lin}}{2005}]{TakeuchiLin2005}
{Takeuchi} T.,  {Lin} D.~N.~C.,  2005, \mn@doi [\apj] {10.1086/428378}, \href
  {https://ui.adsabs.harvard.edu/abs/2005ApJ...623..482T} {623, 482}

\bibitem[\protect\citeauthoryear{{Teague}}{{Teague}}{2019}]{eddy}
{Teague} R.,  2019, \mn@doi [The Journal of Open Source Software]
  {10.21105/joss.01220}, \href
  {https://ui.adsabs.harvard.edu/abs/2019JOSS....4.1220T} {4, 1220}

\bibitem[\protect\citeauthoryear{{Teague} \& {Foreman-Mackey}}{{Teague} \&
  {Foreman-Mackey}}{2018}]{bettermoments}
{Teague} R.,  {Foreman-Mackey} D.,  2018, \mn@doi [Research Notes of the
  American Astronomical Society] {10.3847/2515-5172/aae265}, \href
  {https://ui.adsabs.harvard.edu/abs/2018RNAAS...2c.173T} {2, 173}

\bibitem[\protect\citeauthoryear{{Teague} et~al.,}{{Teague}
  et~al.}{2016}]{Teague2016}
{Teague} R.,  et~al., 2016, \mn@doi [\aap] {10.1051/0004-6361/201628550}, \href
  {https://ui.adsabs.harvard.edu/abs/2016A&A...592A..49T} {592, A49}

\bibitem[\protect\citeauthoryear{{Teague}, {Bae}, {Bergin}, {Birnstiel}  \&
  {Foreman-Mackey}}{{Teague} et~al.}{2018a}]{Teague2018}
{Teague} R.,  {Bae} J.,  {Bergin} E.~A.,  {Birnstiel} T.,   {Foreman-Mackey}
  D.,  2018a, \mn@doi [\apjl] {10.3847/2041-8213/aac6d7}, \href
  {https://ui.adsabs.harvard.edu/abs/2018ApJ...860L..12T} {860, L12}

\bibitem[\protect\citeauthoryear{{Teague} et~al.,}{{Teague}
  et~al.}{2018b}]{Teague2018CS}
{Teague} R.,  et~al., 2018b, \mn@doi [\apj] {10.3847/1538-4357/aad80e}, \href
  {https://ui.adsabs.harvard.edu/abs/2018ApJ...864..133T} {864, 133}

\bibitem[\protect\citeauthoryear{{Teague}, {Bae}, {Birnstiel}  \&
  {Bergin}}{{Teague} et~al.}{2018c}]{Teague2018AS209}
{Teague} R.,  {Bae} J.,  {Birnstiel} T.,   {Bergin} E.~A.,  2018c, \mn@doi
  [\apj] {10.3847/1538-4357/aae836}, \href
  {https://ui.adsabs.harvard.edu/abs/2018ApJ...868..113T} {868, 113}

\bibitem[\protect\citeauthoryear{{Teague}, {Bae}  \& {Bergin}}{{Teague}
  et~al.}{2019}]{Teague2019}
{Teague} R.,  {Bae} J.,   {Bergin} E.~A.,  2019, \mn@doi [\nat]
  {10.1038/s41586-019-1642-0}, \href
  {https://ui.adsabs.harvard.edu/abs/2019Natur.574..378T} {574, 378}

\bibitem[\protect\citeauthoryear{{Testi} et~al.,}{{Testi}
  et~al.}{2014}]{Testi2014}
{Testi} L.,  et~al., 2014, in {Beuther} H.,  {Klessen} R.~S.,  {Dullemond}
  C.~P.,   {Henning} T.,  eds, Protostars and Planets VI. p.~339 (\mn@eprint
  {arXiv} {1402.1354}), \mn@doi{10.2458/azu_uapress_9780816531240-ch015}

\bibitem[\protect\citeauthoryear{{Whipple}}{{Whipple}}{1972}]{1972fpp..conf..211W}
{Whipple} F.~L.,  1972, in {Elvius} A.,  ed., From Plasma to Planet. p.~211

\bibitem[\protect\citeauthoryear{{Youdin} \& {Lithwick}}{{Youdin} \&
  {Lithwick}}{2007}]{Youdin2007}
{Youdin} A.~N.,  {Lithwick} Y.,  2007, \mn@doi [\icarus]
  {10.1016/j.icarus.2007.07.012}, \href
  {https://ui.adsabs.harvard.edu/abs/2007Icar..192..588Y} {192, 588}

\bibitem[\protect\citeauthoryear{{Zhang} et~al.,}{{Zhang}
  et~al.}{2018}]{Zhang2018}
{Zhang} S.,  et~al., 2018, \mn@doi [\apjl] {10.3847/2041-8213/aaf744}, \href
  {https://ui.adsabs.harvard.edu/abs/2018ApJ...869L..47Z} {869, L47}

\bibitem[\protect\citeauthoryear{{van der Marel} et~al.,}{{van der Marel}
  et~al.}{2013}]{vanderMarel2013}
{van der Marel} N.,  et~al., 2013, \mn@doi [Science] {10.1126/science.1236770},
  \href {https://ui.adsabs.harvard.edu/abs/2013Sci...340.1199V} {340, 1199}

\bibitem[\protect\citeauthoryear{{van der Marel}, {Dong}, {di Francesco},
  {Williams}  \& {Tobin}}{{van der Marel} et~al.}{2019}]{vanderMarel2019}
{van der Marel} N.,  {Dong} R.,  {di Francesco} J.,  {Williams} J.~P.,
  {Tobin} J.,  2019, \mn@doi [\apj] {10.3847/1538-4357/aafd31}, \href
  {https://ui.adsabs.harvard.edu/abs/2019ApJ...872..112V} {872, 112}

\bibitem[\protect\citeauthoryear{{van der Plas} et~al.,}{{van der Plas}
  et~al.}{2017}]{vanderPlas2017}
{van der Plas} G.,  et~al., 2017, \mn@doi [\aap] {10.1051/0004-6361/201629523},
  \href {https://ui.adsabs.harvard.edu/abs/2017A&A...597A..32V} {597, A32}

\makeatother
\end{thebibliography}

%%%%%%%%%%%%%%%%%%%%%%%%%%%%%%%%%%%%%%%%%%%%%%%%%%

%%%%%%%%%%%%%%%%% APPENDICES %%%%%%%%%%%%%%%%%%%%%
\appendix

\section{Derivation of the relations for the dust-gas coupling and gas width}
\label{sec:rot_width}
The radial width of a dust ring close to a pressure bump is set by the competition between radial drift (which tends to collect dust at the pressure maximum) and diffusion (that tends to smooth out the ring). Assuming steady state and a zero net dust mass flux, balancing the two terms means solving the following differential equation \citepalias[see e.g.][]{Dullemond2018}:
\begin{equation}
\Sigma_d v_\mathrm{drift} = D_d \frac{\mathrm{d}\Sigma_d}{\mathrm{d}r},
\end{equation}
where $\Sigma_d$ is the dust surface density, $v_\mathrm{drift}$ the radial drift velocity and $D_d$ the diffusion coefficient of the dust. We assume that the diffusion coefficient of the dust is equal to the kinematic viscosity $\nu$, i.e. that the Schmidt number is 1; this is valid for dust with $St \ll 1$ \citep[e.g.,][]{Youdin2007}. Substituting the expression for the radial drift velocity \citep{TakeuchiLin2002}, we obtain
\begin{equation}
\Sigma_d \frac{St}{r} \frac{\mathrm{d}\log p}{\mathrm{d}\log r}= \alpha \frac{\mathrm{d}\Sigma_d}{\mathrm{d}r}.
\label{eq:differential_dust}
\end{equation}

This differential equation contains the logarithmic derivative, that we can put in relation with the rotation curve $v_\phi(r)$ of the gas. The relation between $p(r)$ and $v_\phi(r)$ is given by
\begin{equation}
v_\phi(r) = v_K(r) +\frac{1}{2}\frac{c_s^2}{v_K}\frac{d\log p(r)}{d\log r}.
\end{equation}
Calling $\delta v_\phi\equiv v_\phi-v_K$ (the deviation from Keplerian) we can write
\begin{equation}
\frac{\delta v_\phi}{v_K} = \frac{1}{2}\frac{c_s^2}{v_K^2}\frac{d\log p(r)}{d\log r}.
\label{eq:vphi_order}
\end{equation}
In the razor-thin case $p_{2\mathrm{D}}=c_s^2 \Sigma$ and it is useful to use this to rewrite this expression as:
\begin{equation}
\frac{\delta v_\phi}{v_K} = \frac{1}{2}\frac{c_s^2}{v_K^2} \left(\frac{d\log \Sigma}{d\log r}+ \frac{d\log c_s^2}{d\log r} \right).
\label{eq:vphi_order_sigma}
\end{equation}
We will not use this expression in the context of the razor-thin analysis, but introducing it is nevertheless useful for the analysis in appendix \ref{sec:vertical_structure}. \autoref{eq:vphi_order} can be used to find the logarithmic derivative of the pressure:
\begin{equation}
\frac{d\log p(r)}{d\log r} = 2 \frac{v_k^2}{c_s^2} \frac{\delta v_\phi}{v_k}.
\label{eq:dlogpdlogr}
\end{equation}
Substituting it into \autoref{eq:differential_dust} leads to the following differential equation for the dust structure:
\begin{equation}
2 \Sigma_d \frac{St}{r} \frac{v^2_k}{c_s^2} \frac{\delta v_\phi}{v_k} = \alpha \frac{\mathrm{d}\Sigma_d}{\mathrm{d}r}.
\label{eq:differential_dust_vphi}
\end{equation}
In principle, we could solve this equation for the dust structure given the $\delta v_\phi/v_k$ measured by the observations. However, this is not straightforward (for example, we already stated that we do not know the true keplerian value). A more robust approach is to employ the derivative of the velocity measured close to a pressure maximum $r_0$, which is equivalent to Taylor-expanding the rotation curve (or the logarithmic derivative of the pressure profile):
\begin{equation}
\frac{\delta v_\phi}{v_k} = \left. \frac{\mathrm{d}}{\mathrm{d}r} \left( \frac{\delta v_\phi}{v_k} \right)  \right|_{r_0} (r-r_0) + O [(r-r_0)^2]
\end{equation}
since by construction the rotation curve vanishes at $r_0$. Because $r_0$ is a maximum, we also deduce that $\mathrm{d}/\mathrm{d}r (\delta v_\phi/v_k)<0$ in the vicinity of $r_0$.

With this approximation, and to first order in $(r-r_0)$, \autoref{eq:differential_dust_vphi} becomes
\begin{equation}
2 \Sigma_d St  \frac{v^2_k}{c_s^2} \frac{\delta v_\phi}{v_k} \frac{(r-r_0)}{r_0} = \alpha \frac{\mathrm{d}\Sigma_d}{\mathrm{d}r}.
\end{equation}
The solution of this differential equation is a Gaussian:
\begin{equation}
\Sigma_d = \Sigma_{d0} \exp \left[ -\frac{(r-r_0)^2}{2w_d^2}\right],
\end{equation}
where we have introduced the width $w_d$, which is given by
\begin{equation}
w_d^2=-\frac{1}{2} \frac{\alpha}{St} \frac{c_s^2 r_0}{v_k^2}  \left[ \frac{\mathrm{d}}{\mathrm{d}r} \left. \left( \frac{\delta v_\phi}{v_k}\right) \right|_{r_0} \right]^{-1}.
\label{eq:width_dust}
\end{equation}
Recalling that $\mathrm{d}/\mathrm{d}r (\delta v_\phi/v_k)<0$, we can see that $w_d^2$ is as expected a positive quantity. Inverting the last equation we finally get to the final expression that links $\alpha/St$ with the observables:
\begin{equation}
\frac{\alpha}{St} = - \frac{2 w_d^2}{r_0} \frac{v^2_k}{c_s^2} \frac{\mathrm{d}}{\mathrm{d}r} \left. \left( \frac{\delta v_\phi}{v_k}\right) \right|_{r_0}.
\label{eq:alpha_St}
\end{equation}

\subsection{The gas structure}
\label{sec:width_gas}

The first-order expansion of the rotation curve also allows us to write the pressure profile in the proximity of the pressure maximum. Using the expansion to first order of the rotation curve we can rewrite \autoref{eq:dlogpdlogr} as:
\begin{equation}
\frac{d\log p(r)}{d\log r} = 2 \frac{v_k^2}{c_s^2} \left. \frac{\mathrm{d}}{\mathrm{d}r} \left( \frac{\delta v_\phi}{v_k} \right)  \right|_{r_0} (r-r_0). 
\end{equation} 
Integrating this equation we obtain that
\begin{equation}
p(r) = p_0 \exp \left[ - \frac{B}{2r_0} (r-r_0)^2 \right] = p_0 \exp \left[ - \frac{ (r-r_0)^2}{2w_g^2} \right],
\end{equation}
where we have called $w_g$ the width of the gas, which is linked to the observables as follows:
\begin{equation}
w_g=\sqrt{-\frac{1}{2}\frac{c_s^2}{v_K^2}r_0\left[\frac{d}{dr}\left(\frac{\delta v_\phi}{v_K}\right)\right]^{-1}}.
\label{eq:width_gas}
\end{equation}
The expansion to first order we have done in this paper is therefore equivalent to assume that the gas pressure profile is a Gaussian. The same quantities that we use to estimate $\alpha/St$ can also be used to measure the width of this Gaussian.

\section{Vertical structure}
\label{sec:vertical_structure}
%This derivation largely follows \citet{TakeuchiLin2002} but generalises it to the case of a generic vertical temperature gradient. 
We now drop the assumption of a razor-thin disc and consider the disc vertical structure. Force balance in the radial direction reads
\begin{equation}
\frac{v^2_\phi}{r}=\frac{GMr}{\left(r^2+z^2\right)^{3/2}}+\frac{1}{\rho} \frac{\partial p}{\partial r}.
\end{equation}
We now define $v_K^2 = GMr^2/(r^2+z^2)^{3/2}$, i.e. the Keplerian velocity at height $z$, and use that $p= c_s^2 \rho$. As before, we introduce $\delta v_\phi\equiv v_\phi-v_K$ and use these quantities to rewrite this expression as
\begin{equation}
\frac{\delta v_\phi}{v_K}=\frac{1}{2} \frac{c_s^2}{v^2_k} \left(\frac{\partial \log \rho}{\partial \log r}+ \frac{\partial \log c_s^2}{\partial \log r}\right),
\label{eq:master}
\end{equation}
where to simplify to notation we have not marked explicitly the dependence on $z$ of the various quantities; we will follow this convention also in the next equations, except than when it is needed to resolve ambiguities. %Note that here we defined $v_K^2 = GMr^2/(r^2+z^2)^{3/2}$, i.e. the Keplerian velocity at height $z$. In the same fashion, also the sound speed is evaluated at the height $z$. 
Note that, while in the razor-thin case we directly prescribed a structure in the gas pressure, it is now necessary to distinguish between density and temperature because they have a different dependence on the vertical coordinate. We now focus on the term $\partial \log \rho/\partial \log r$.  Without loss of generality, $\rho(r,z)=\rho_0 (r) f_H(r,z)$, where $f_H(r,z)$ is such that $f_H(r,z=0)=1$. We assume that most of the mass is concentrated close to the midplane, so that $\rho_0 \propto \Sigma/H$ neglecting the details of the vertical structure (which is valid for realistic temperature profiles, e.g. \citealt{Flock2013}), where $H=c_\mathrm{s,midplane}/\Omega_k$ is the gas scale-height in the midplane.  With this notation
\begin{equation}
\frac{\partial \log \rho}{\partial \log r}= 
\frac{\partial \log \Sigma}{\partial \log r}-\frac{\partial \log H}{\partial \log r}+\frac{\partial \log f_H(r,z)}{\partial \log r}.
\end{equation}
The density and sound speed in the vertical direction must satisfy the vertical hydrostatic equilibrium:
\begin{equation}
\frac{1}{\rho} \frac{\mathrm{d}p}{\mathrm{d}z} = \frac{1}{\rho} \frac{\mathrm{d}(\rho c_s^2)}{\mathrm{d}z} = \frac{G M z}{(r^z+z^2)^{3/2}}.
\end{equation}
If $c_s(z)$ is known, this equation is separable and can be directly integrated; the formal solution, as can be verified by substituting it in the previous expression, reads
\begin{equation}
\rho = \rho_0 \frac{c_\mathrm{s,midplane}^2}{c_s^2(z)} \exp \left( - \int_0^z \frac{\Omega_k^2 z' \mathrm{d}z'}{c_s^2(z')} \right),
\end{equation}
which specifies $f_H(r,z)$. We further assume that the sound speed $c_s$ varies in the following way with height:
\begin{equation}
c_s^2(r,z)=c_\mathrm{s,midplane}^2(r) f_c(r,z)=c_\mathrm{s,midplane}^2(r) g(z/H(r)),
\end{equation}
i.e., that the increase in temperature depends only on $z/H$. It is then natural to introduce the dimensionless variable $x=z/H$. A commonly used functional shape for $g$, first proposed by \citet{Dartois2003}, is
\begin{equation}
g(x)=1+(\theta - 1)\sin^4\left( \frac{\pi x}{x_\mathrm{trans}} \right),
\end{equation}
where $\theta$ is a free parameter specifying the ratio between the temperatures in the midplane and in the atmosphere, while $x_\mathrm{trans}$ is the vertical coordinate (in units of the scale-height) where the temperature transitions to the value in the atmosphere.

After some algebra, we get to the final expression:
\begin{align}
\frac{\delta v_\phi}{v_K}= \frac{1}{2} \frac{c_s^2}{v^2_k} & \left\{ \frac{\partial \log \Sigma}{\partial \log r} + \frac{\partial \log c_{\mathrm{s,midplane}}^2}{\partial \log r}  \right. + \nonumber \\
& %\qquad \qquad \qquad \qquad 
\left.  \frac{\partial \log H}{\partial \log r} \left[ \int_0^{z/H} \left( \frac{2x}{g} - \frac{x^2 g'}{g^2} \right) \mathrm{d}x - 1 \right]  \right\},
\label{eq:delta_vphi_2d}
\end{align}
where $g'=\mathrm{d}g/\mathrm{d}x$ and the integral (a dimensionless number) can easily be evaluated numerically for a given choice of $g$. Note that this expression correctly reduces to the isothermal limit given by \citet{TakeuchiLin2002}, in which case $g \equiv 1$ and $g'=0$.

\begin{figure}
    \centering
    \includegraphics[width=\columnwidth]{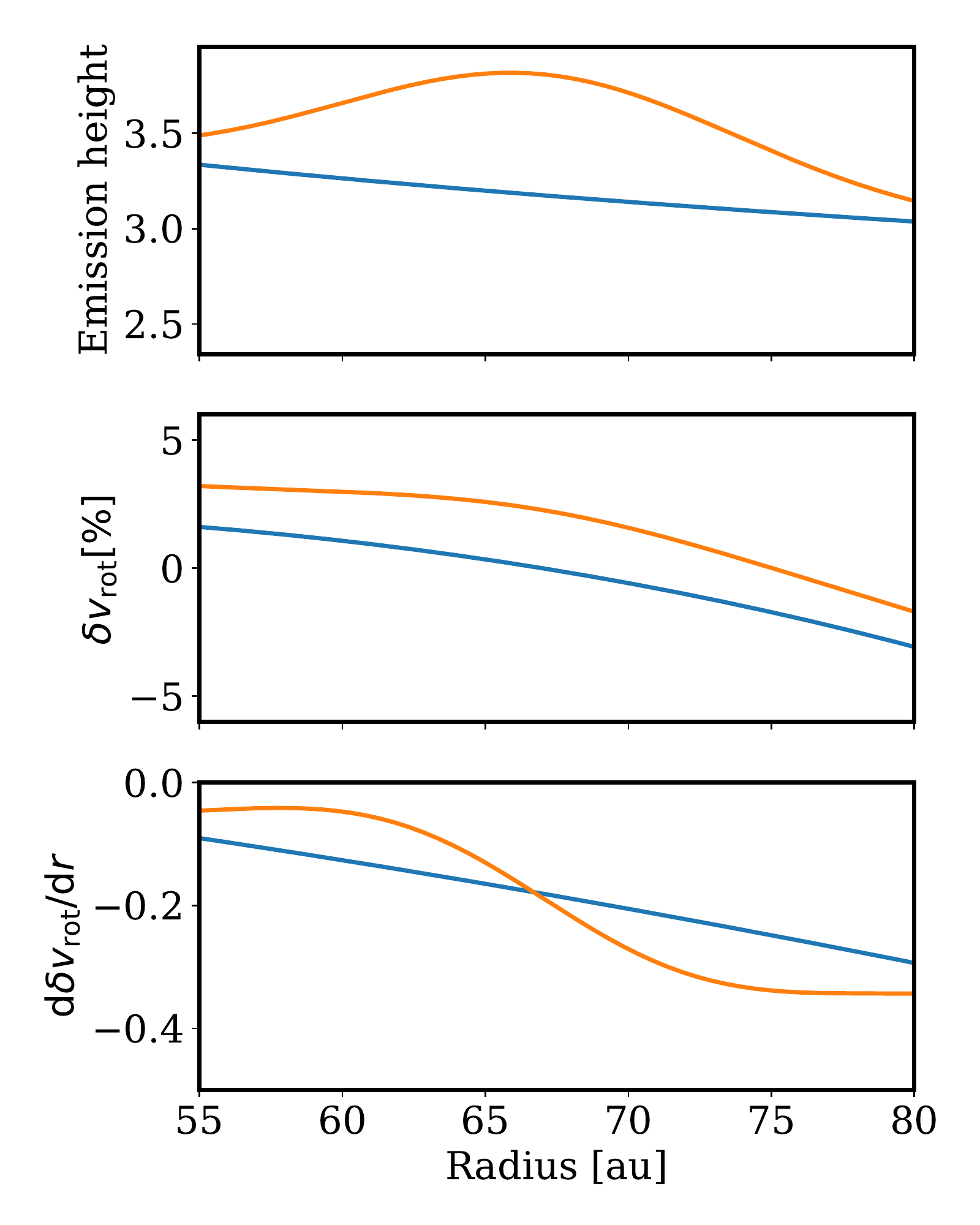}
    \caption{Illustration of the effect of the disc vertical structure, and in particular of a varying height of the emission surface. The blue line depicts the case of a constant $z/r$ of the emission surface (note that this means that height of the emission surface, measured in scale-heights, slightly decreases with radius because the scale-height increases with radius), while in the orange line we consider a local increase in the height of the emission surface. This leads to a morphological change in the shape of the rotation curve, but it does not affect the average value.}
    \label{fig:perturb}
\end{figure}

This formula allows us to study the validity of \autoref{eq:alpha_St} and \autoref{eq:width_gas} in comparison with \autoref{eq:vphi_order_sigma}. The most obvious change is that the temperature to use is not the one in the midplane, but the one at a height $z$, because the term $c_s^2$ in front of the parenthesis is now evaluated at a height $z$. The other difference is that this equation contains additional terms inside the parenthesis; for clarity we reported these terms on the second line. To study the impact of these terms, it is worth remembering that in this paper we use the slope of the rotation curve, i.e. the derivative of \autoref{eq:delta_vphi_2d}. Although the additional terms present in this equation might potentially be non-negligible, they are constant with radius as long as a) the temperature can be described as a power-law and b) the height of the emission surface (measured in scale-heights) does not change. Therefore, these terms will not introduce biases as long as those two conditions are satisfied, although they do introduce a constant offset in the perturbation of the rotation curve from Keplerian\footnote{Recall that in our methodology we do not know anyway the true Keplerian value. The additional terms in \autoref{eq:delta_vphi_2d} are the very reason why in general we expect an offset from the Keplerian value.}. Regarding a), we note that the brightness temperature emission profile of $^{12}$CO is relatively smooth (see Fig. 7 of \citealt{Isella2018} and Fig. 5 of \citealt{Guzman2018}), which justifies our assumption of neglecting a radial temperature gradient on the same spatial scales of the pressure bump. We note that the $^{12}$CO-derived temperature is the one at the emission surface; there is limited information for the temperature in the mid-plane, but at least for HD163296, the method of \citet{Dullemond2019} also finds a smooth temperature profile. Regarding point b), it is reasonable to expect that close to a pressure maximum, due to the increase in surface density, the height of the emission surface might have a local increase. Because in general the integral in \autoref{eq:delta_vphi_2d} increases with $z$, this produces a local perturbation in the rotation curve around the local maximum that is always super-Keplerian and has a maximum at the pressure maximum. It follows that its derivative changes sign at the pressure maximum. We illustrate this graphically in \autoref{fig:perturb}. In the figure, we have modelled this local increase as a Gaussian with the same width as the perturbation in surface density and assumed an amplitude of the perturbation of 20 per cent; we used parameters corresponding to B67 in HD163296 \citepalias[mid-plane temperature of 30K, following][and temperature in the upper layers of 80K]{Dullemond2018}. The figure shows that the perturbation induced by a variation of the height of the emission surface is \textit{morphologically} different from the one induced by the pressure maximum and therefore should not bias our determination of the width. Note that in principle the perturbation could have a rather high amplitude in the value of the slope (see bottom panel); however, it does not affect the average value because the effect has two different signs on the two sides of the pressure maximum. Finally, we note that this effect also tends to shift towards the outside the apparent location of the pressure maximum derived from the rotation curve.

\bsp	% typesetting comment
\label{lastpage}
\end{document}